\documentclass[aps,nofootinbib,prd,eqsecnum,showpacs,showkeys,preprintnumbers,nofootinbib,altaffilletter]{revtex4}
\usepackage{graphicx}
\usepackage{color}
\usepackage{euscript,amssymb}
\usepackage{amsfonts}
\usepackage{amsmath}
\usepackage{amssymb}
\usepackage{graphicx}
\usepackage{euscript}
\usepackage{hyperref}
\usepackage{epstopdf}
\usepackage{url}
\usepackage{footnote}
\usepackage{mathtools}

\begin{document}

\title{Can $f(R)$ gravity contribute to (dark) radiation?}

\author{Jo\~{a}o Morais}
\email{jviegas001@ikasle.ehu.eus}
\affiliation{Department of Theoretical Physics,\\University of the Basque Country UPV/EHU,\\ P.O. Box 644, 48080 Bilbao, Spain,}

\author{Mariam Bouhmadi-L\'{o}pez}
\email{mbl@ubi.pt - on leave of absence from UPV/EHU and IKERBASQUE}
\affiliation{ Departamento de F\'isica, Universidade da Beira Interior, 6200 Covilh\~a, Portugal,}
\affiliation{Centro de Matem\'atica e Aplica\c{c}\~oes da Universidade da Beira Interior (CMA-UBI), 6200 Covilh\~a, Portugal,}
\affiliation{Department of Theoretical Physics,\\University of the Basque Country UPV/EHU,\\ P.O. Box 644, 48080 Bilbao, Spain,}
\affiliation{
IKERBASQUE, Basque Foundation for Science, 48011, Bilbao, Spain.}

\author{Salvatore Capozziello}
\email{capozziello@na.infn.it}
\affiliation{ Dipartimento di Fisica, Universit\`a di Napoli "Federico II, Compl. Univ. di Monte Sant'Angelo, Via Cinthia 9, I-80126 Napoli, Italy,}
\affiliation{Istituto Nazionale di Fisica Nucleare (INFN) Sez. di Napoli, Compl. Univ. di Monte Sant'Angelo, Via Cinthia 9, I-80126 Napoli, Italy,}
 \affiliation{Gran Sasso Science Institute (INFN), Viale F. Crispi 7, I-67100 L'Aquila, Italy.}

\begin{abstract}
We discuss the possibility that suitable modifications of gravity could account for some amount of the radiation we observe today, in addition to the possibility of explaining the present speed up of the universe. We start introducing and reviewing cosmological reconstruction methods for metric $f(R)$ theories of gravity that can be considered as one of the straightforward modifications of Einstein's gravity as soon as $f(R)\neq R$. We then take into account two possible $f(R)$ models which could give rise to (dark) radiation. Constraints on the models are found by using the Planck Collaboration 2015 data within a cosmographic approach and by obtaining the matter power spectrum of those models. The conclusion is that $f(R)$ gravity can only contribute minimally to the (dark) radiation to avoid departures from the observed matter power spectrum at the smallest scales (of the order of $0.01$Mpc$^{-1}$), i.e., precisely those scales that exited the horizon at the radiation dominated epoch. This result could strongly contribute to select reliable $f(R)$ models.
\end{abstract}

\date{\today}



\maketitle


\section{Introduction}

\emph{Precision Cosmology} is giving new perspectives in setting and defining self-consistent cosmological models whose reliability is based on a good deal of high quality data \cite{Riess:1998cb,Perlmutter:1998np,Knop:2003iy,Tonry:2003zg,Barris:2003dq,Riess:2004nr}. In fact, considering the today state of art, several cosmological models are capable of grossly reproducing the cosmic expansion history since their bounds match cosmic data in a wide range of redshift \cite{Copeland2006,Sahni:2006pa}. This situation started at the end of last century. Before, cosmologists assumed that the cosmic energy budget was given essentially by pressureless matter density (dust) sourcing a decelerated Hubble flow. After 1998, observations pointed out that such a Hubble flow is undergoing an accelerated expansion. In other words, it was evident that observational results could not be interpreted only adopting baryons and dark matter as sources in the Einstein-Friedmann cosmological equations. As a consequence, the Cosmological Standard Model had to be modified, including, at least, the cosmological constant $\Lambda$, within the energy momentum tensor \cite{Sahni:1999gb,Saini:1999ba}. In this sense, $\Lambda$ represented the first straightforward explanation of the current observed acceleration \cite{Carroll:1991mt,Carroll:2000fy,Alam:2004jy,Li2011}.
 
However, the puzzle was and is to interpret the physical nature of the cosmological constant. Several hypotheses came out but, up to now, none is fully satisfactory. For example, $\Lambda$ can be related to the non-zero gravitational vacuum energy and can be framed in the context of quantum field theory in curved spacetime \cite{Birrell:1982ix,Parker:2011}. Despite of this fundamental physics explanation, theoretical predictions and cosmological observations show a difference of a huge amount of orders of magnitude, leading to the dramatic \emph{cosmological constant problem} that escapes any standard physics explanation \cite{Weinberg1989}. Besides, comparing today the densities of matter and $\Lambda$, the result is extremely close to each other in order of magnitudes ($\Omega_M\sim 0.25\div 0.3$ and $\Omega_{\Lambda}\sim0.7\div 0.75$), leading to an extremely fine-tuning called \emph{coincidence problem} \cite{Peebles:2002gy}. It consists with the fact that there is no reasons to expect that matter and $\Lambda$ densities have to be comparable at present time: standard matter evolves as the universe expands, while $\Lambda$ should be constant at all stages of the evolution. In other words, the difference between the two fluids should be huge today despite the observations coming from Precision Cosmology. Alternatively, one can suppose that the fluid sourcing the cosmic acceleration could not have a constant density with the same order of magnitude along the whole cosmic expansion \cite{Padmanabhan:2002ji,Sahni:2002fz,Melchiorri:2002ux,Carroll:2003st,Alam:2003sc,Alam:2003fg,Carroll:2004hc,Bean:2008ac,Silvestri:2009hh,Basilakos:2010fb,Grande:2011xf,Basilakos:2012ra,Lima:2012mu,Perico:2013mna,BouhmadiLopez:2013pn,Basilakos:2014tha}.

In turn, any standard fluid lying in the Zeldovich interval $(0\leq w \leq 1)$ cannot work since, as soon as it is inserted into the cosmological equations, gives rise to a decelerated behavior. This means that we have to resort to exotic fluids. However, up to now, we have no final experimental evidence for the existence of these fluids at fundamental level. In general, such a dynamics is addressed under the standard of \emph{dark energy} \cite{Huterer:1998qv,Durrer:2007re,Tsujikawa:2010zza}. Furthermore, comparing both late and early cosmic phases of the evolution led to the question to search for suitable scalar fields, capable of describing de-Sitter-like behaviors. In principle, such an approach would be useful to relate early inflation and today observed acceleration into a single description 
\cite{Wetterich:1987fk,Wetterich:1987fm,Peebles:1987ek,Hebecker:2000zb,Pavlov:2013nra}.

A natural way to address the problem is in terms of curvature invariants and geometric corrections \cite{Capozziello:2002rd,Capozziello:2003tk,Carroll:2003wy,Nojiri:2003ft,Allemandi:2005qs,Nojiri:2006gh,Capozziello:2006dj,Starobinsky:2007hu,Bertolami:2007gv,Boehmer:2007fh,Boehmer:2007tr,Sotiriou:2007zu}. The approach is based on results at ultraviolet scales, where additional curvature and non-local terms emerge into the Hilbert-Einstein Lagrangian as a consequence of the formulation of quantum field theory in curved spacetime \cite{Birrell:1982ix,Brandenberger:1984cz,Parker:2011}. In other words, {\it scalar fields} can be derived from geometry providing viable interpretations of dark energy and inflation as geometric effects at large scales (infrared) and small scales (ultraviolet), respectively \cite{Nojiri:2006ri,Capozziello:2007ec,DeFelice:2010aj,Sotiriou2010,Nojiri:2010wj,Capozziello:2011et,Nojiri2014,Starobinsky1980,Capozziello:2002rd,Capozziello:2005ku,delaCruzDombriz:2006fj,Nojiri:2006be,Dunsby:2010wg,Carloni:2010ph}.

This geometric view is an extension of General Relativity (GR) aimed to cure shortcomings at low and high energy scales \cite{Bassett:2005xm,Fay:2007gg,Tsujikawa:2007xu}, without introducing dark components \cite{Capozziello:2011et}. In general, dynamics should be sketched as follows. The universe starts at very high curvature regimes, then the curvature decreases and GR is restored at intermediate scales. Finally, infra-red corrections start to work at very large scales. Saying it differently, the huge initial curvature gives rise to acceleration (inflation), the decreasing of curvature allows deceleration (matter dominated era) and then, further decreasing, allows the growth of sub-dominant terms and the transition from deceleration to acceleration. This phenomenology fixes the critical points of the cosmic evolution and could give rise to structure formation \cite{Capozziello:2014zda}. 
In summary, the early-time and the late-time cosmic speed-up could be related to the fact that curvature corrections provide significative consequences at large and small energy regimes \cite{Carroll:2004de,Nesseris:2005ur,DeFelice:2006pg,Bean:2006up,Durrer:2008in,Bamba:2008ut,Bamba:2008hq,Bamba:2009uf,Bamba:2012ky}. 

 In other words, one may recover a dynamical effective $\Lambda$, taking care of the cosmological constant problem and assuming it as a limiting case of a general dynamics governed by curvature terms \cite{Fay:2007uy,Nesseris:2013jea}. In this picture, the dark energy is today mimicked by $\Lambda$ that appears only as a zero-order term of more general Extended Theories of Gravity \cite{Capozziello:2011et}, as, for example, $f(R)$ gravity.
This is clear if one assumes the effective gravitational action coming from some fundamental theory. In fact, we do not need to add any cosmological constant {\it by hand} but we need to reproduce it in some low-energy limit or symmetric (vacuum) configuration. For example, $\Lambda$ can be derived from $f(R)$ gravity where the cosmological constant appears as a sort of eigenvalue \cite{Capozziello:2007gm}. A {\it smoking gun} for this approach seems to be given by the primordial perturbation spectrum as discussed in \cite{Vilenkin:1985md,Starobinsky:2007hu,BouhmadiLopez:2012qp,Bouhmadi-Lopez:2013jfm,Basilakos:2014moa,Ade:2015rim}. 

Despite of these interesting features, theories with higher-order curvature terms show some defects and shortcomings that have to be fixed \cite{Nojiri:2010wj,Amendola:2006kh,Amendola:2006we,Appleby:2007vb,Appleby:2008tv,Kusakabe:2015yaa}. In any case, modified gravities constitute a useful approach able to comply observational data with cosmic phenomenology \cite{Hu:2001bc,Faraoni:2005vk,Faraoni:2006hx,Carroll:2006jn,Sawicki:2007tf,Hu:2007pj,Hu:2007nk,Bertolami:2008ab,Bertolami:2008zh,Appleby:2008tv,Hu:2013aqa,Dossett:2014oia,Raveri:2014cka,Sotiriou2010,Basilakos:2013nfa,Basilakos:2014moa,Bertolami:2013uwl}. Furthermore, modified gravities can provide also self-consistent explanations for dark matter \cite{Boehmer:2007kx,Capozziello:2008ima,Capozziello:2012ie}. In this sense, the whole dark sector could be encompassed in a comprehensive geometric picture.

From a more formal point of view, modified gravity theories can be formulated with different approaches, e.g. 
 metric, Palatini and purely affine. The differences depend, essentially, on the fact that the causal and geodesic structures coincide (adopting the 
 Levi-Civita connection) or do not coincide (as in the Palatini formalism which results in a bi-metric theory \cite{Olmo:2011uz,Bouhmadi-Lopez:2013nma,Bouhmadi-Lopez:2014tna,Bouhmadi-Lopez:2014jfa}). Furthermore, such theories can be formulated in a purely affine fashion considering only connection as a fundamental variable \cite{Poplawski:2007mq}.
 
In this paper, we will focus on $f(R)$ metric theories of gravity being the straightforward extension of GR that provides a ground for the inflationary and the late-time acceleration \cite{Liddle:1988tb,Barrow:1988xh,Liddle:1993fq,Liddle:2000,DeFelice:2010aj,Nojiri:2010wj,Capozziello:2011et}. In particular, we shall investigate the possibility that $f(R)$ models, beside dark energy, could contribute as well to some amount of the cosmic radiation giving rise to a sort of {\it dark radiation}. This feature could be relevant in order to find out a signature for higher-order gravity.

The concept of a dark radiation, an unknown relativistic component that only interacts gravitationally with normal matter, was introduced to indicate the excess of radiation in cosmological observation and that cannot be explained by photons or the three families of neutrinos from the Standard Model \cite{Ade:2015xua}. This excess of radiation is parametrized as extra relativistic degrees of freedom, leading to a deviation from the Standard Model prediction of $N_{eff}= 3.046$ \cite{Mangano:2005cc}. Historically, the sterile neutrino models were initially employed to explain the nature of dark radiation, \cite{Archidiacono:2011gq,Abazajian:2012ys}. However inconsistencies with observations led to the introduction of different candidates to explain such a component, such as {\em dark electromagnetism} mediated by a {\em dark photon} \cite{Ackerman:mha}, thermal axions from Quantum Chromodynamics \cite{Archidiacono:2011gq}, or even the result of dark matter particles decay \cite{Hasenkamp:2012ii,Menestrina:2011mz}. Although the recent Planck Collaboration results give a value of $N_{eff}$ very much consistent with the predictions of the Standard Model, an excess of radiation is still allowed. In particular, seems to be consistent with a higher value of the current Hubble parameter \cite{Ade:2015xua}.

The paper is organized as follows. In Sec.~\ref{Sec2}, we introduce cosmological reconstruction methods for $f(R)$ theories and apply them to some models. In Sec.~\ref{Sec3}, we present two homogeneous and isotropic models, where an $f(R)$ description interpolates between two cosmological behaviours corresponding to a radiation dominated phase and a cosmological constant phase. In the first model the radiation is completely dark, while in the second one it is only partially dark. In Sec.~\ref{Sec4}, we review the cosmographic approach for $f(R)$ gravity and write down the relations between the cosmographic and the cosmological parameters for the models introduced in Sec.~\ref{Sec3}. In Sec.~\ref{Sec5}, we review the first order perturbations theory of $f(R)$ and write down the formulas for the evolution of the perturbed quantities. Then, the matter power spectrum is obtained for the models introduced in Sec.~\ref{Sec3} without imposing any sort of approximation on the perturbed equations, i.e. a full numerical integration is carried out. Discussion and conclusions are drawn in Sec.~\ref{Sec6}. The main result of the paper is that $f(R)$ gravity minimally contributes to the cosmic radiation background.

\section{A reconstruction method for $f(R)$ gravity starting from the equation of state}
\label{Sec2}

We start constructing a metric $f(R)$ model such that for a homogeneous and isotropic universe, the cosmological expansion is equivalent to that of a relativistic model filled by a fluid with an equation of state $p=p(\rho)$ (see Refs.~\cite{Nojiri:2006ri,Capozziello:2007ec,DeFelice:2010aj,Nojiri:2010wj,Capozziello:2011et,Sotiriou2010,Nojiri2014} for reviews on $f(R)$-gravity and \cite{Capozziello:2002rd,Capozziello:2005ku,delaCruzDombriz:2006fj,Nojiri:2006be,Dunsby:2010wg,Carloni:2010ph,Nojiri:2010wj} for earlier works on reconstruction methods in $f(R)$ gravity). We will refer to two cosmological expansions as equivalent if the geometrical quantities in both expansions are the same, i.e. $H$, $\dot H$, $R$ and $\dot R$ are equal.

In general, the cosmological expansion of a relativistic universe is described by the Friedmann and the Raychaudhuri equations
\begin{align}
	\label{Friedmannn1}
	H^2&=\frac13 \rho,\\
	\dot H&=-\frac12 (p+\rho),
\end{align}
where we have set $\kappa_4^2=8\pi G=1$, $G$ is the gravitational constant, and a dot stands for a derivative with respect to the cosmic time. Therefore, the scalar curvature in terms of the matter content reads
\begin{align}
	R&=12H^2+6\dot H\nonumber \\
	&=\rho-3p.
	\label{curvature}
\end{align}
On the one hand, using the conservation and Friedmann equations we obtain
\begin{align}
	\label{rho_p_dev}
	\dot\rho = -\sqrt{3\rho}(p+\rho), \quad \dot p=-\sqrt{3\rho}(p+\rho)\frac{dp}{d\rho},
\end{align}
where we have assumed $p=p(\rho)$. The last two equations, \eqref{curvature} and \eqref{rho_p_dev}, imply
\begin{align}
	\dot R= -\sqrt{3\rho}(p+\rho)\left(1-3\frac{dp}{d\rho}\right).
	\label{Rdot}
\end{align}

On the other hand, in the $f(R)$ metric scenario, the gravitational action reads
\begin{align}
	\label{f(R)_action}
	\mathcal{S} = \frac{1}{2}\int\sqrt{-g}\textrm{d}^4\textbf{x}\,f(R).
\end{align}
As mentioned before, we consider $\kappa_4^2=1$. The $f(R)$ action \eqref{f(R)_action} leads to the modified Friedmann equation, which in the Jordan frame, reads
\begin{align}
	3H^2\frac{df}{dR}=\frac12\left(R\frac{df}{dR}-f\right)-3H\dot R \frac{d^2f}{dR^2} + \rho_m.
	\label{Friedmannn2}
\end{align}
A relativistic model which satisfies Eqs.~(\ref{Friedmannn1})-(\ref{Rdot}) can be described through an $f(R)$ expansion as long as there is an $f(R)$ solution to
\begin{align}
	3\rho(\rho+p)\left(1-3\frac{dp}{d\rho}\right)\frac{d^2f}{dR^2}-\frac12(\rho+3p)\frac{df}{dR}-\frac12f=0.
	\label{eqdiff1}
\end{align}
The previous equation is obtained from Eq.~\eqref{Friedmannn2}, with $\rho_m=0$, and assuming that the relativistic and the $f(R)$ cosmological expansions are equivalent. Now, as we are assuming that the pressure is determined by the energy density $\rho$, we can conclude that the scalar curvature is exclusively determined by $\rho$. A similar argument applies to $f(R)$ which it is exclusively determined by $\rho$. It is therefore helpful to rewrite Eq.~(\ref{eqdiff1}) as
\begin{align}
	3\rho(\rho+p)\left(1-3p'\right)f''+\left[9\rho(p+\rho)p''-\frac12(\rho+3p)(1-3p')\right]f'-\frac12\left(1-3p'\right)^2f=0,
	\label{eqdiff2}
\end{align}
where a prime stands for a derivative with respect to the energy density $\rho$. In summary, any homogeneous and isotropic relativistic model fuelled by a perfect fluid with energy density $\rho$ and pressure $p(\rho)$ can be described by an $f(R)$ model as long as $f$ satisfies Eq.~(\ref{eqdiff2}).
This approach is more evident in the Palatini formulation where the form of $f(R)$ function is directly given by the matter density (see \cite{Olmo:2011uz,Capozziello:2007ec} for details).

In this work, we use a reconstruction method for $f(R)$ gravity where we assume an effective equation of state for the effective energy density of $f(R)$. Notice that this is simply a rephrasing of our previous results, as we assume that the conservation of the energy momentum tensor holds or equivalently that the Bianchi identities are fulfilled. Once we have $p(\rho)$, we can define $R(\rho)$ and solve the Friedmannn equation for $f(\rho)$. As long as the equation can be solved analytically, and we can invert the relation $R(\rho)$, this method has the advantage of providing us with an expression for $f(R)$. However, the $f(R)$ solutions obtained in this way can have very complicated expressions which may not be easily interpreted in a physical sense \cite{Capozziello:2008qc,{BouhmadiLopez:2012qp}}.
Furthermore, by assuming an equation of state of the kind $p=p(\rho)$ we are restricting the degrees of freedom of our gravitational theory. In general, we would have
\begin{align}
	\label{p_of_derivs}
	p = p\left(\rho,\dot\rho,\ddot\rho,...\right).
\end{align}
Thus, when we define $R(\rho)$, and consequently $\rho(R)$, we are ignoring the dependence of $p$ on $\dot\rho$ and the other derivatives of the density (see Eq.~\eqref{p_of_derivs}). An alternative method to study the dynamics of $f(R)$ gravity that takes into account the dependence in Eq.~\eqref{p_of_derivs} is by doing a dynamical system analysis of the Friedmannn and Raychaudhury equations \cite{Carloni:2004kp,Amendola:2006we,Carloni:2007eu,Amendola:2007nt,Tsujikawa:2007xu,Carloni:2009nc,Abdelwahab:2011dk}. However, this kind of analysis usually requires the class of $f(R)$ functions to be defined \textit{a priori} in order to get a closed set of equations. In summary, any of the methods mentioned above have their own advantages and disadvantages.

%
%

\subsection{Some examples}

We apply the reconstruction procedure introduced previously to some simple but still interesting models.

%
%

\subsubsection{The cosmological constant case}

We start considering the simplest case where the perfect fluid satisfies a cosmological constant equation of state, i.e. $p=-\rho$. Then the constraint \eqref{eqdiff2} reads \cite{Dunsby:2010wg}
\begin{align}
	\rho f'-2f=0,
\end{align}
hence, $f\propto \rho^2$ which implies
\begin{align}
	\label{CosmConst_solution}
	f=C\,R^2,\quad C=\rm{Const}.
\end{align}
The behaviour $p=-\rho$ is special in the sense that the dynamical variables (except the scale factor) are static as $\dot R=\dot H =0$. If we consider a de Sitter solution in vacuum, we find that the Friedmann equation reduces to \cite{Faraoni:2005vk,Barrow:1983rx} \begin{align}
	R_{dS}\left(\frac{df}{dR}\right)_{{dS}}-2f\left(R_{dS}\right)=0,
\end{align}
where a $dS$ subscript indicates evaluation of a quantity at the de Sitter solution. This is an algebraic equation that dictates the possible de Sitter points for each $f(R)$ function. The solution $f=C\,R^2$ is special in the sense that it satisfies the de Sitter condition for every $R$, while for other functions we only obtain a discrete set of de Sitter points.

%
%

\subsubsection{The constant parameter case for the equation of state}

We assume that $p=w\rho$ where $w\neq 1/3$ and constant. Then Eq. \eqref{eqdiff2} reduces to
\begin{align}
	3(1+w)\rho^2 f''-\frac12(1+3w)\rho f'-\frac12 (1-3w)f=0.
\end{align}
It can be shown that the solution of the previous equation is of the form $f\propto \rho^{\beta_{\pm}}$ where
\begin{align}
	\label{constant_w_exponents}
	\beta_\pm=\frac12\left\{1+\frac{1+3w}{6(1+w)}\pm\sqrt{\frac{2(1-3w)}{3(1+w)}+\left[1+\frac{1+3w}{6(1+w)}\right]^2}\right\}.
\end{align}
The argument on the square root in Eq.~\eqref{constant_w_exponents} vanishes when $w=-(13\pm4\sqrt6)/3$, being negative inside the interval defined by those points and positive otherwise. Thus, for $w>-1$, the exponents $\beta_\pm$ are always real valued.
As the scalar curvature is a linear function of the energy density, we conclude that \cite{Capozziello:2002rd,Dunsby:2010wg}
\begin{align}
	\label{f_sol_w_const}
	f(R)=C_+R^{\beta_+} + C_-R^{\beta_-}, \quad C_\pm=\rm{const}.
\end{align}

Here we have fixed the dependence of $p(\rho)$, and consequently of $R(\rho)$, which allowed us to obtain a linear differential equation for $f(R(\rho))$ (see Eq.~\eqref{eqdiff2}). We can try to pursue the opposite approach, i.e., fixing $f(R(\rho))$ or $f(R)$, as given for example in Eq.~\eqref{f_sol_w_const}, and look for the appropriate equation of state $p=w\rho$. More precisely, by plugging a given $f(R(\rho))$ or $f(R)$ into Eq.~\eqref{eqdiff1}, we would obtain a first order non-linear differential equation for $p(\rho)$. When following this approach for $f(R)$ given in Eq.~\eqref{f_sol_w_const}, we obtain that $p=w\rho$ is a solution of the non linear differential equation fulfilled by $p(\rho)$ but there is a further solution which for briefness we have omitted.

We now focus our attention on the case $p=w\rho$ with $w=1/3$. In Eq.~\eqref{eqdiff2} the term $p''$ vanishes for a constant $w$. If we divide the equation by $(1-3p')$, and then set $p'=w=1/3$ and $p=\rho/3$, we obtain the differential equation
\begin{align}
	4\rho^2f'' - \rho f' =0.
\end{align}
The general solution of this equation is the linear combination
\begin{align}
	\label{}
	f = C_1 + C_2\rho^{5/4}.
\end{align}
These exponents correspond to the limits of Eq.~\eqref{constant_w_exponents} as $w$ approaches $1/3$. In this particular case $R=0$. This result is strictly related to the conformal invariance of radiation solutions. For a detailed discussion on this topic see \cite{Capozziello:2010zz}.

%
%

\subsection{A Modified Generalised Chaplygin Gas}

Let us now consider an $f(R)$ model which mimics the behaviour of a homogeneous and isotropic universe filled by a modified generalised Chaplygin gas (mGCG) \cite{Kamenshchik:2001cp,Bento:2002ps,Benaoum:2002zs,GonzalezDiaz:2002hr,Bento:2003dj,BouhmadiLopez:2004me,BouhmadiLopez:2004mp,Chimento:2004bv,Chimento:2005au,BouhmadiLopez:2007qb,BouhmadiLopez:2006fu,BouhmadiLopez:2009hv,BouhmadiLopez:2011kw,BouhmadiLopez:2012by} whose equation of state reads 
\begin{align}
	\label{mGCG_eos}
	p_{ch} = \beta\rho_{ch} - (1+\beta)\frac{A}{\rho_{ch}^\alpha}.
\end{align}
 For earlier attempts of describing the GCG in $f(R)$ gravity see Ref.~\cite{Capozziello:2005ku}.
The conservation of the energy momentum tensor implies that the energy density of the mGCG scales as
\begin{align}
	\label{mGCG_of_a}
	\rho_{ch}(a) = \rho_{ch,0}\left[A_s + (1-A_s)\left(\frac{a_0}{a}\right)^{3\xi}\right]^{\frac{1}{1+\alpha}}.
\end{align}
Here, we have defined the energy density of the mGCG at the present time, $\rho_{ch,0}$, the factor $\xi$ is defined as $\xi\equiv(1+\beta)(1+\alpha)$, and $A_s$ is a dimensionless constant such that $A_s\equiv A/\rho_{ch,0}^{1+\alpha}$. From now on, a subscript $0$ stands for a quantity evaluated at the present time, except if stated otherwise. We restrict our analysis to the case $0<A_s<1$, which is the simplest condition required for a mGCG to fuel a late time or an early acceleration epoch in the universe. If we define the scale factor $a_*$
\begin{align}
	a_* = \left(\frac{1-A_s}{A_s}\right)^{\frac{1}{3\xi}}a_0,
\end{align}
and rewrite Eq.~\eqref{mGCG_of_a} as
\begin{align}
	\rho_{ch}(a) = \rho_{ch,0}A_s^{\frac{1}{1+\alpha}}\left[1 + \left(\frac{a_*}{a}\right)^{3\xi}\right]^{\frac{1}{1+\alpha}},
\end{align}
we can identify two distinct regimes of the mGCG model in Eq.~\eqref{mGCG_of_a}. In fact, for $\xi>0$, we have that 
\begin{align}
	\label{mGCG_behave}
	\rho_{ch} &\approx 
	\begin{dcases}
		\rho_{ch,0}A_s^{\frac{1}{1+\alpha}}\left(\frac{a_*}{a}\right)^{3(1+\beta)}, & \mbox{if }a\ll a_*, \\
		\rho_{dS}, & \mbox{if } a\gg a_*,
	\end{dcases}
\end{align}
where $\rho_{dS}\equiv\rho_{ch,0}A_s^{1/(1+\alpha)}$, while, for $\xi<0$, we obtain the same behavior but in an inverted chronological order
\begin{align}
	\rho_{ch} &\approx 
	\begin{dcases}
		\rho_{dS}, & \mbox{if }a\ll a_*, \\
		\rho_{ch,0}A_s^{\frac{1}{1+\alpha}}\left(\frac{a_*}{a}\right)^{3(1+\beta)}, & \mbox{if } a\gg a_*.
	\end{dcases}
\end{align}

We next find the appropriate $f(R)$ function which mimics the behavior of a mGCG for an empty universe. We substitute Eq.~\eqref{mGCG_eos} into Eq.~(\ref{eqdiff2}) in order to obtain the second order linear differential equation fulfilled by $f(\rho)$
\begin{align}
	\label{eqdiff_mGCG1}
	&6\rho^2(1+\beta)\left[1 - \left(\frac{\rho_{dS}}{\rho}\right)^{1+\alpha}\right] \left[1-3\beta-3\alpha\left(\frac{\rho_{dS}}{\rho}\right)^{1+\alpha}\right]f'' \nonumber\\
	&- \rho\left\{ \left[1-3\beta+3(1+\beta)\left(\frac{\rho_{dS}}{\rho}\right)^{1+\alpha}\right] \left[1-3\beta-3\alpha(1+\beta)\left(\frac{\rho_{dS}}{\rho}\right)^{1+\alpha}\right] + 18\alpha\xi(1+\beta)\left(\frac{\rho_{dS}}{\rho}\right)^{1+\alpha}\right\}f' \nonumber\\
	&-\left[1-3\beta-3\alpha(1+\beta)\left(\frac{\rho_{dS}}{\rho}\right)^{1+\alpha}\right]^2f=0.
\end{align}
For later calculations, it is helpful to introduce the dimensionless variable $x$, defined as
\begin{align}
	\label{def_x}
	x = \left(\frac{\rho_{dS}}{\rho}\right)^{1+\alpha},
\end{align}
which at present is such that $x_0=A_s$. The variable $x$ is finite and restricted to be within the interval $(0,1)$: $x$ approaches unity when the mGCG is near the de Sitter regime, and vanishes asymptotically when the energy density of the mGCG scales as $a^{-3(1+\beta)}$. Eq.~\eqref{eqdiff_mGCG1} can be rewritten in terms of the variable $x$ as follows
\begin{align}
	\label{eqdiff_mGCG2}
	\frac{d^2f}{dx^2}
	&+ \left[\frac{6\xi + 9\beta+7}{6\xi}\frac{1}{x} + \frac{1}{3\xi}\frac{1}{x-1} - \frac{1}{x - \frac{1-3\beta}{3\alpha(1+\beta)}}\right]\frac{df}{dx} + \frac{1}{2(1+\alpha)^2}\left[-\alpha + \frac{1-3\beta}{3(1+\beta)}\frac{1}{x}\right]\frac{f}{x(x-1)}=0.
\end{align}
The general solution of this equation can be expressed as a linear combination of the two solutions
\begin{align}
	\label{hyper_sol}
	f_1(x) =&
		\frac{(3 {\beta} +5-\lambda_\beta)(3\xi+1)
		+ 12(1+\beta)}{3}
	\left( x-1 \right)
	{x}^{ 1 - \frac {9\beta + 7 + \lambda_\beta}{ 12\xi}}
	{\mbox{F}\left[
		1 + \frac{3 {\beta} +5-\lambda_\beta}{ 12\xi},
		2 - \frac{3\beta + 1 + \lambda_\beta}{ 12\xi} ;
		2 - \frac {\lambda_\beta}{ 6\xi};
		x
	\right]} \nonumber\\
	&+\frac{6\xi - \lambda_\beta}{6} \Big[
			\left(9\beta+11 - \lambda_\beta\right)x
		-
			\left(9\beta+7 - \lambda_\beta\right)
	\Big]
	{x}^{-\frac {\lambda_\beta + 9\beta + 7}{ 12\xi }}{\mbox{F}\left[
		\frac {3\beta + 5 - \lambda_\beta}{ 12\xi},
		1 - \frac {3\beta + 1 + \lambda_\beta}{ 12\xi};
		1 - \frac {\lambda_\beta}{ 6\xi};\,x
	\right]},
	\nonumber\\
	f_2(x) =& 
	\frac{
		(3 {\beta} +5 + \lambda_\beta)(3\xi+1)
		+12(1+\beta)}{3}
	\left( x-1 \right)
	{x}^{1 + \frac {\lambda_\beta - 9\beta -7}{ 12\xi }} {\mbox{F} \left[
		1 + \frac {3\beta + 5 + \lambda_\beta}{ 12\xi},
		2 - \frac {3\beta +1 - \lambda_\beta}{ 12\xi};
		2 + \frac {\lambda_\beta}{ 6\xi};
		x
	\right]} \nonumber\\
	&+ \frac{6\xi + \lambda_\beta}{6}
			\Big[(9\beta + 11 + \lambda_\beta)x-(9\beta + 7 + \lambda_\beta)
	\Big]
	{x}^{\frac {\lambda_\beta -9\beta - 7}{12 \xi}}{\mbox{F} \left[
		\frac{3\beta + 5 + \lambda_\beta}{ 12\xi},
		1 - \frac {3\beta + 1 - \lambda_\beta}{ 12\xi};
		1 + \frac {\lambda_\beta}{ 6\xi};
		x
	\right]}.
\end{align}
where $\textrm{F}[b,c;d;x]$ is a hypergeometric function \cite{Abramowitz1965,Olver2010} and $\lambda_\beta\equiv\sqrt {9{\beta}^{2}+78\beta+73}$.

%
%

\section{The background $f(R)$ models}
\label{Sec3}

\subsection{A model for dark radiation}

We remind the goal of this work: Can a modified theory of gravity like $f(R)$ account not only for the current acceleration of the universe but also contribute by some amount to the (dark) radiation of the universe? As a first step to address this question, we will use the model building introduced in the previous section. More precisely, we will consider the solution \eqref{hyper_sol} with $\beta=1/3$ and $1+\alpha>0$, i.e. an $f(R)$ model that interpolates between an early radiation epoch and a late time de Sitter phase.

In this case, the differential equation that the $f(R)$ function, mimicking a mGCG (with $\beta=1/3$), must fulfill is much simpler than the one given in Eq.~\eqref{eqdiff_mGCG2}, as it corresponds to a hypergeometric differential equation \cite{Abramowitz1965,Olver2010}
\begin{align}
	\label{eqdiff_B=1/3}
	x(1-x)\frac{d^2f}{dx^2} - \frac{6x-5}{4(1+\alpha)}\frac{df}{dx} + \frac{\alpha}{2(1+\alpha)^2}f=0.
\end{align}
This equation admits, as general solution around the point $x=0$, the linear combination \cite{Abramowitz1965,Olver2010}
\footnote{This solution is valid in the range $(0,1)$ as long as $5/(4+4\alpha)$ is not an integer \cite{Abramowitz1965,Olver2010}. We have disregarded the case $5/(4+4\alpha)$ equal to an integer as it is extremely fine tuned.}
\begin{align}
	\label{solution_f1_f2}
	f(x) =& C_1f_1(x) + C_2f_2(x) \nonumber\\
	=& C_1\textrm{F}
	\left[
		\frac{1}{2(1+\alpha)},
		-1+\frac{1}{1+\alpha};
		\frac{5}{4(1+\alpha)};
		x
	\right] 
	+ C_2x^{1-\frac{5}{4(1+\alpha)}}
	\textrm{F}
	\left[
		-\frac{1}{4(1+\alpha)},
		1-\frac{3}{4(1+\alpha)};
		2-\frac{5}{4(1+\alpha)};
		x
	\right].
\end{align}
Additionally, we can invert the relation $R(\rho)$, by combining Eqs.~\eqref{curvature} and \eqref{mGCG_eos}, and use Eq.~\eqref{def_x} to write the variable $x$ in terms of the scalar curvature
\begin{align}
	\label{x_of_R}
	x(R) = \left(\frac{R}{R_{dS}}\right)^{1 + \frac{1}{\alpha}}.
	\end{align}
Here $R_{dS}\equiv4\rho_{dS}$ is the asymptotic value of the scalar curvature when the mGCG enters the de Sitter expansion. Finally, the solution $f(R)$ reads
\begin{align}
	\label{solution_f(R)_B=1/3}
	f(R) =& C_1\textrm{F}
	\left[
		\frac{1}{2(1+\alpha)},
		-1+\frac{1}{1+\alpha};
		\frac{5}{4(1+\alpha)};
		\left(\frac{R}{R_{dS}}\right)^{1 + \frac{1}{\alpha}}
	\right] \nonumber\\
	&+ C_2\left(\frac{R}{R_{dS}}\right)^{1 - \frac{1}{4\alpha}}
	\textrm{F}
	\left[
		-\frac{1}{4(1+\alpha)},
		1-\frac{3}{4(1+\alpha)};
		2-\frac{5}{4(1+\alpha)};
		\left(\frac{R}{R_{dS}}\right)^{1 + \frac{1}{\alpha}}
	\right].
\end{align}
Setting $\beta=1/3$ in the solutions \eqref{hyper_sol} and making use of the relations between contiguous hypergeometric functions \cite{Abramowitz1965,Olver2010}, we arrive at the same result of Eq.~\eqref{solution_f(R)_B=1/3}.

In Eq.~\eqref{solution_f(R)_B=1/3}, we express the general function $f(R)$ that is compatible with the mGCG model (Eq.~\eqref{mGCG_eos} with $\beta=1/3$). However, the linear coefficients $C_1$ and $C_2$ need yet to be specified in order for the function $f(R)$ to be physically meaningful. This will be done imposing the physical constraints that the $R$ derivatives of the function $f$ must fulfill \cite{Hu:2007nk,Starobinsky:2007hu}
\begin{enumerate}
	\item	Gravity is attractive since the Big Bang nucleosynthesis, therefore
		\begin{align}
			f_R(x)>0 ~~~~,\forall x.
		\end{align}
		Here, we introduce the notation $f_R\equiv df/dR$. From now on, $R$ subscripts indicate derivatives with respect to the scalar curvature.
	\item	The effective gravitational constant at the present time must match $G$, i.e. the standard gravitational constant, therefore
		\begin{align}
			f_R(x_0) =1.
		\end{align}
	\item The existence of stable de Sitter solutions \cite{Faraoni:2005ie,Faraoni:2006hx} imposes that:
 	\begin{align}
		m^2_{\textrm{eff}}=\frac{{f_R}^2-2f\,f_{RR}}{f_R\,f_{RR}}>0.
	\end{align}
	Notice that this condition is much weaker than the previous two because: (i) the early ``de Sitter-like'' inflationary phase of the universe is unstable in the sense that the universe must exit it at the reheating epoch; and (ii) while the most likely scenario for the future of our universe is a de Sitter-like phase, we cannot guarantee it for sure as this conclusion is based on our current knowledge about the matter content of the universe and on the supposition that it will remain so for ever.
 \item The scalaron is not a tachyon \cite{Starobinsky:2007hu}
	 \begin{align}
			f_{RR}(x)>0 ~~~~,\forall x.
	\end{align}
	This condition is intrinsically related to the Dolgov-Kawasaki instability which was first discovered for the model $f(R)=R-\mu^4/R$ \cite{Dolgov:2003px} and later generalized for any $f(R)$ function \cite{Faraoni:2006sy}.
\end{enumerate}

In appendix~\ref{app_B}, we give the expressions of $f_{1R}$, $f_{1RR}$, $f_{2R}$ and $f_{2RR}$ in terms of the variable $x$. Using those results, we obtain the values of $\alpha$ for which the functions $f_1$ and $f_2$, as well as their $R$ derivatives, are well defined at the points $x=0$ and $x=1$. Our results are presented in Table~\ref{conv_alpha}.
\footnote{We remind that the function F$[b,c;d;x]$, defined by the hypergeometric series \cite{Abramowitz1965,Olver2010}, converges for any value $|z|\leq1$, whenever $b+c-d<0$. However, for $0\leq b+c-d<1$, the series does not converge at $z=1$, and for $1<b+c-d$ it is divergent at $|z|=1$. We notice as well that F$[b,c;d;0]=1$.}

\begin{table}
	\begin{tabular}[t]{| l | c | c || l | c | c |}
		\hline
		 		& $x=0$					& $x=1$ 						& 			& $x=0$						& $x=1$\\ \hline\hline
		$f_1$ 	& $\forall\alpha$ 			& $\alpha<-1~~\textrm{or}~~-3/4<\alpha$	& $f_2$		& $\alpha<-1~~ \textrm{or} ~~1/4<\alpha$	& $\alpha<-1~~\textrm{or}~~-3/4<\alpha$ \\
		$f_{1R}$	& $-1<\alpha$				& $\alpha<-1$					& $f_{2R}$		& $\alpha<-1$					& $\alpha<-1$	\\	
		$f_{1RR}$	& $-1<\alpha<1$			& $-5/4<\alpha<-1$				& $f_{2RR}$	& $\nexists\alpha$				& $-5/4<\alpha<-1$	\\
		\hline
	\end{tabular}
	\caption[Convergence intervals for the parameter $\alpha$]{\label{conv_alpha}Intervals of the parameter $\alpha$ for which the solutions $f_1$ and $f_2$ (given in Eq.~\eqref{solution_f1_f2}) and their first $R-$derivatives are finite and well-defined.}
\end{table}

\begin{figure}[t]
	\includegraphics[width=\textwidth]{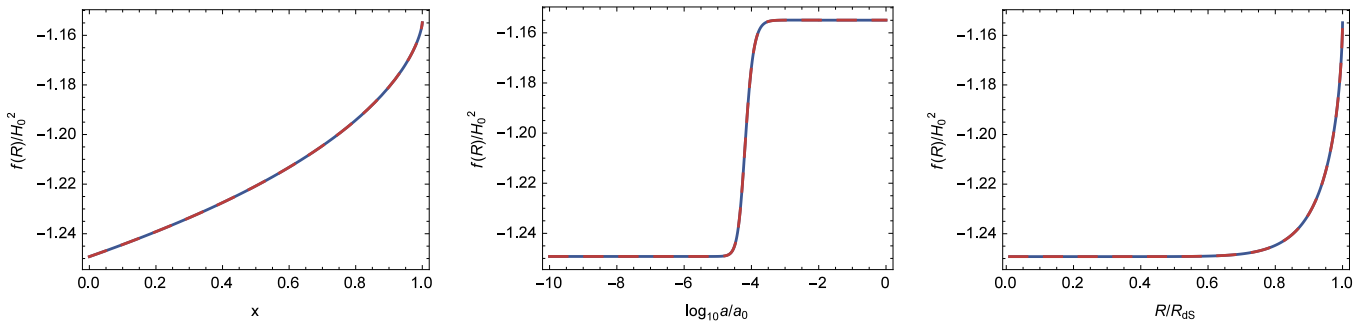}
	\caption{\label{f(R)_vacuum}The analytical solution $f(R)$ (blue curve) versus the numerical solution of $f(R)$ (red dashed curve) in the vacuum case; i.e. the analytical solution $C_1f_1(R)$ and the numerical solution of Eq.~(\ref{Friedmannn2}) for the effective matter content corresponding to the mGCG with $\beta=\frac13$ (cf. Eq.~\eqref{solution_f1_f2}). The numerical integration has been carried out imposing that at present $f(R)$ and $df/dR$ take the values corresponding to the analytical solution. These plots show clearly that our analytical solution is correct. In this case, the values of the parameters $(\alpha,A_s)$ are $\alpha=0.104$ and $A_s=A_s^{(m)}=0.999948$, as given in Table~\ref{back_para_tab}.}
\end{figure}

To choose the value of the linear coefficients $C_1$ and $C_2$, we take into account some additional physical conditions, namely, the radiation like expansion must occur in the past, i.e. $1+\alpha>0$, and the scalar curvature vanishes asymptotically during that period, i.e. $\alpha>0$. Additionally, we require that the function $f$ and its two derivatives $f_R$ and $f_{RR}$ are finite in the past. With these considerations in mind, and given the results of Table~\ref{conv_alpha}, we set $C_2=0$ (i.e., $f(x)=C_1f_1(x)$) and restrict our analysis to values of $\alpha$ such that $0<\alpha<1$. 
From Eqs.~\eqref{f1R_B=1/3} and \eqref{f1RR_B=1/3}, we find that $f_{1R}$ and $f_{1RR}$ are negative for all values of $x\in[0,1]$, if $\alpha>0$. Therefore, the requirements $f_R>0$ and $f_{RR}>0$ are automatically satisfied if we set $C_1<0$. The condition $f_R(x_0)=1$ fixes the value of $C_1$ as
\begin{align}
	C_1(\alpha,x_0) = \frac{1}{f_{1R}(x_0)}
	= -10\rho_{dS}\frac{x_0^{-\frac{1}{1+\alpha}}}{
	\textrm{F}
	\left[
		1+\frac{1}{2(1+\alpha)},
		\frac{1}{1+\alpha};
		1+\frac{5}{4(1+\alpha)};
		x_0
	\right]}.
\end{align}
Notice that with this definition and for $0<\alpha<1$, we have $C_1<0$ for all values of $x_0$, therefore, we have that $f_R>0$ and $f_{RR}>0$. The solution $f(R)=C_1f_1(R)$ is plotted in Fig.~\ref{f(R)_vacuum}. Finally, we can conclude that $m^2_{\textrm{eff}}$ is positive for any value of $\alpha$, i.e. the late time de Sitter phase is stable.

\subsection{Including dust-like matter}
\label{Sec3B}
\begin{figure}[ht]
	\includegraphics[width=\textwidth]{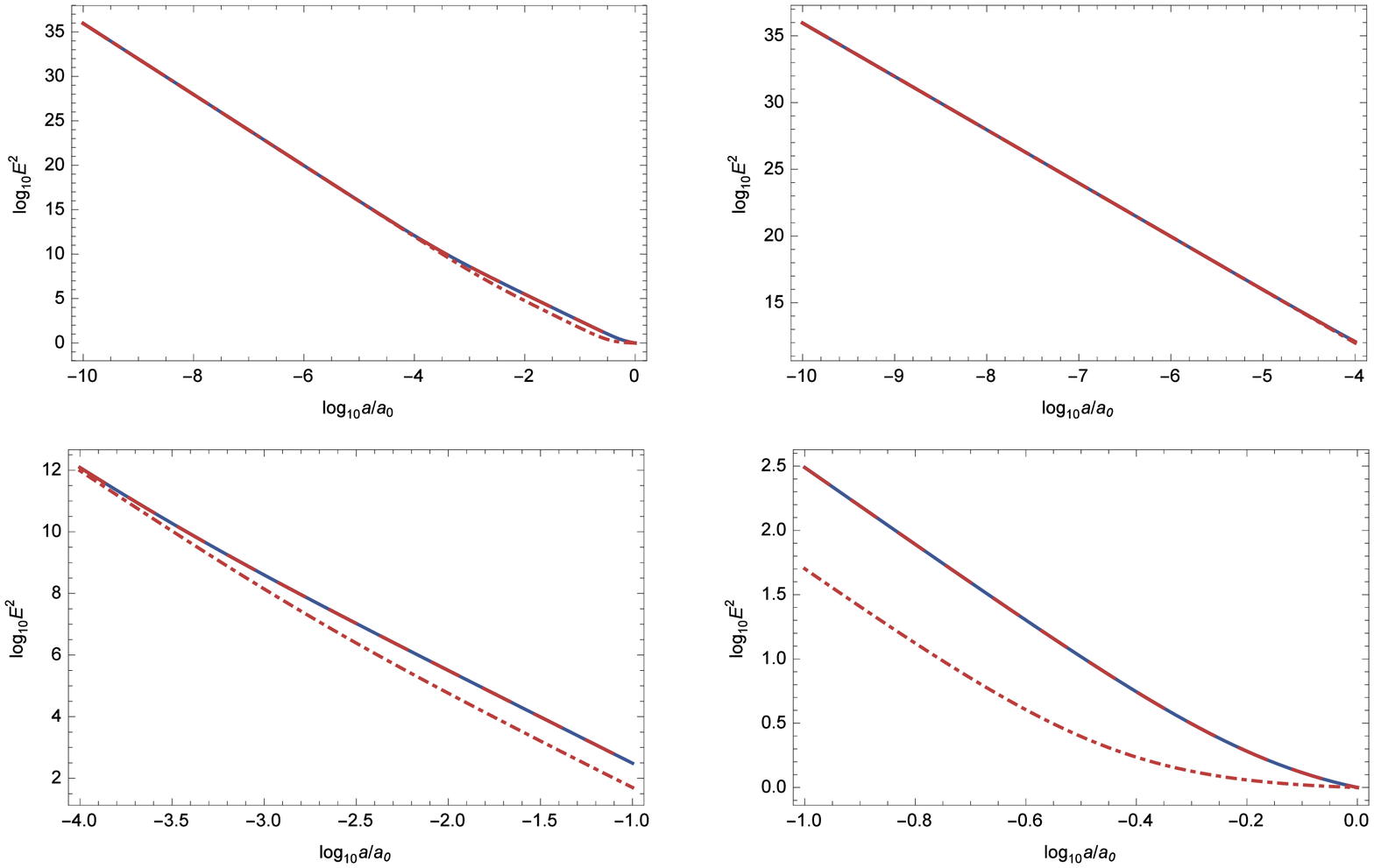}
	\caption{These figures show the behaviour of the square of the dimensionless Hubble parameter $E^2=(H/H_0)^2$ for the $\Lambda$CDM model (blue curve), the mGCG model where a cold dark matter is present (red dashed curve) or not (red dot-dashed curve). The top left figure shows the evolution since the radiation epoch until the present time. The top right figure, bottom left, and bottom right figures show the evolution during the radiation epoch, the matter epoch, and the beginning of the dark energy dominated epoch, respectively. The values of the model parameters are given in Table~\ref{back_para_tab}.
	 }
\label{E2_evolution}
\end{figure}

In the previous subsection we have obtained the physical $f(R)$ function that can mimic a mGCG given in Eq.~\eqref{mGCG_of_a} with $\beta=1/3$ ($0<\alpha<1$ and $0<A_s<1$). As a next step, we consider the presence of dust-like matter, i.e. dark and baryonic matter, so that the background evolves as
\begin{align}
	\label{background}
	3H^2 = \rho_{m,0}\left(\frac{a_0}{a}\right)^{3} + \rho_{ch,0}\left[A_s + (1-A_s)\left(\frac{a_0}{a}\right)^{4(1+\alpha)}\right]^{\frac{1}{1+\alpha}},
\end{align}
where $\rho_{m,0}$ is the present energy density for dark and baryonic matter. In this model, the universe undergoes three distinct epochs. During the first epoch, the universe is dominated by the mGCG which behaves as radiation with the density parameter
\begin{align}
	\Omega_{ch}(a) \approx \frac{\rho_{ch,0}}{3H_0^2}(1-A_s)^{\frac{1}{1+\alpha}}\left(\frac{a_0}{a}\right)^{4}.
\end{align}
Afterwards, the energy density of the dust component surpasses that of the mGCG and the universe enters the matter dominated epoch. Finally, as the expansion continues the behaviour of the mGCG switches smoothly to that of a cosmological constant with
\begin{align}
	\Omega_{ch}(a) \approx \Omega_{dS} = \frac{\rho_{ch,0}}{3H_0^2}A_s^{\frac{1}{1+\alpha}},
\end{align}
where the matter dominated epoch ends and the universe starts to accelerate.

The parameter that defines the equation of state of the mGCG, $w_{ch}$, reaches zero at some point during the evolution of the universe. If the condition $w_{ch}\approx0$ holds for long enough then the mGCG model could also account for the dark matter content of the universe, with a relative density of
\begin{align}
	\Omega_{chm}(a) &\approx \frac{\rho_{ch,0}}{3H_0^2}\left(\frac{1-A_s}{3A_s}\right)^{\frac{3}{4(1+\alpha)}}\left(4A_s\right)^{\frac{1}{1+\alpha}}\left(\frac{a_0}{a}\right)^{3},
	~~~~~~~~\textrm{for } \; w_{ch}\approx0.
\end{align}
In Fig.~\ref{E2_evolution}, we compare the evolution of the square of the dimensionless Hubble parameter, $E^2=(H/H_0)^2$, for the $\Lambda$CDM model with a radiation content (blue curve), the mGCG model with a baryonic and dark matter content (red dashed curve) and the mGCG model with a baryonic content (red dot-dashed curve). Using the results of Ref.~\cite{Ade:2015xua}, we set the values of the cosmological parameters expressed in Table~\ref{back_para_tab}, which we use in Fig.~\ref{E2_evolution} and the subsequent figures.
\begin{table}
	\begin{tabular}[t]{| l | c | c |}
		\hline
		 	~~$\Omega_{m,0}$~~			& $0.3065$ 				& density parameter of dark and baryonic matter\\
		 	~~$\Omega_{b,0}$~~			& $0.0485$				&density parameter of baryonic content\\
		 	~~$\Omega_{\Lambda}$~~		& $0.6935$				&density parameter of the cosmological constant\\
		 	~~$\Omega_{r,0}$~~			& $\Omega_{m,0}/(1+z_{eq})$	&density parameter of radiation content\\
		 	~~$z_{eq}$~~				& $3361$				&redshift at matter radiation equilibrium\\
		 	~~$\alpha$~~				& $0.104$				&$\alpha$ parameter of the mGCG model\\
		 	~~$\Omega_{ch,0}^{(m)}$~~		& $1- \Omega_{m,0}$		&density parameter of the mGCG in the presence of dark and baryonic matter\\
		 	~~$\Omega_{ch,0}^{(b)}$~~		& $1- \Omega_{b,0}$		&density parameter of the mGCG in the presence of baryonic matter\\
		 	~~$A_s^{(m)}$~~			&~~$1-\left(\Omega_{r,0}/\Omega_{ch,0}^{(m)}\right)^{1+\alpha}$~~	&$A_s$ parameter of the mGCG model in the presence of dark and baryonic matter\\
		 	~~$A_s^{(b)}$~~				&~~$1-\left(\Omega_{r,0}/\Omega_{ch,0}^{(b)}\right)^{1+\alpha}$~~	&$A_s$ parameter of the mGCG model in the presence of baryonic matter\\
		\hline
	\end{tabular}
	\caption[Parameters of the background]{\label{back_para_tab}Values of the background parameters used in the numerical analysis.}
\end{table}

We find that, during the matter era, the mGCG model with only a baryonic content deviates too much from the $\Lambda$CDM evolution and as such is cosmologically unviable. This reflects the fact that in the mGCG model with $\beta=1/3$, the condition $w_{ch}\approx0$ is not satisfied for a long enough period of time for the model to account for the dark matter content. Therefore, we have to incorporate a dark matter component as explicitly shown in Eq.~\eqref{background}.

\begin{figure}[t]
	\includegraphics[width=\textwidth]{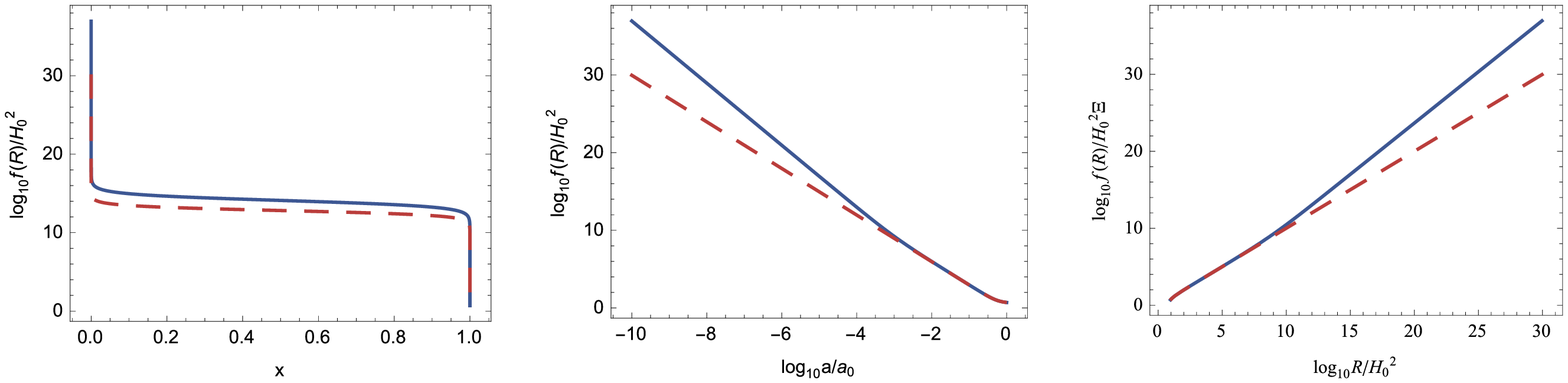}
	\caption{\label{numerical_darkmatter}The numerical solution of $f(R)$ (blue curve), that accounts for the matter content corresponding to the mGCG with $\beta=\frac13$ in the presence of a dark matter and baryonic matter content, versus the action $R-2\Lambda$ (red dashed curve). The numerical integration has been carried out imposing that at present $f(R_0)=R_0-2\Lambda$ and $f_R(R_0)=1$. The values of the parameters are given in Table~\ref{back_para_tab} .
	}
\end{figure}
\begin{figure}[t]
	\includegraphics[width=\textwidth]{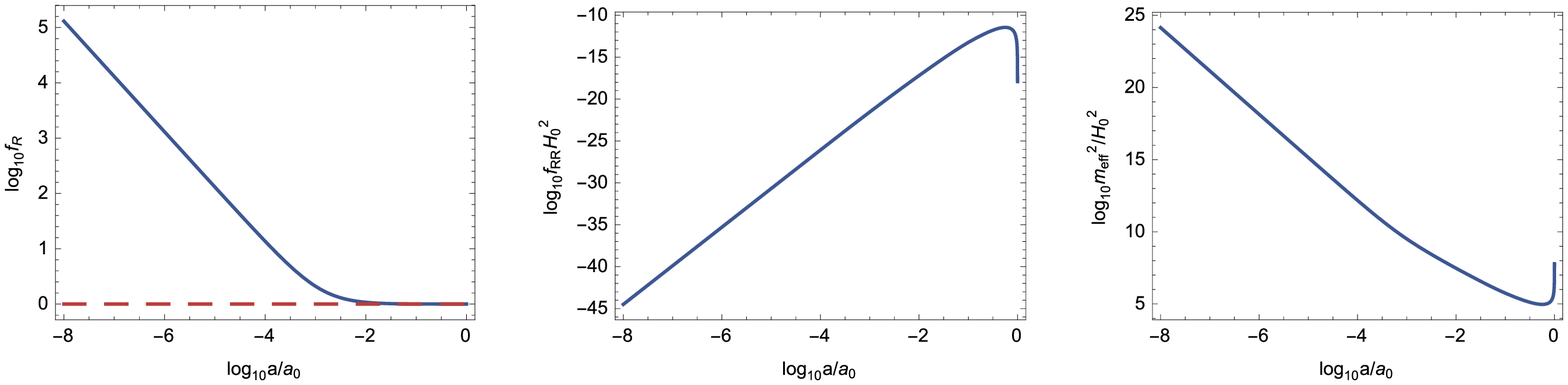}
	\caption{\label{derivs_darkmatter}The evolution of: $f_R$ (left panel); $f_{RR}$ (middle panel); and $m_{eff}^2$ (right panel); for the numerical solution in Fig.~\ref{numerical_darkmatter} (blue curves). A dashed red curve indicates the value of the quantity in GR. While $f_R$ remains positive throughout the entire evolution, $f_{RR}$ and $m_{eff}^2$ become negative in the future, when $a> a_0$, which will give rise to instabilities in the future. We use the values in Table~\ref{back_para_tab} .
	}
\end{figure}

To obtain an $f(R)$ function that is compatible with a mGCG in the presence of dust-like matter, we can no longer use Eq.~\eqref{eqdiff2}.
Instead, we use Eq.~\eqref{Friedmannn2} with $\rho_m$ corresponding to dark and baryonic matter to obtain a second order differential equation for $f(R(a))$. In this case, while finding an analytical solution for $f(R)$ is impossible, we can integrate the equation numerically with proper boundary conditions ($f(a_0)\approx R(a_0)-2\Lambda$ and $f_R(a_0)=1$) to obtain $f(a)$. Those boundary conditions imply that the deviation from GR at the present time is not extreme. The solution obtained is plotted in Fig.~\ref{numerical_darkmatter}, where we compare it with the Einstein-Hilbert action $R-2\Lambda$. During the late time evolution, the $f(R)$ function closely mimics the Einstein-Hilbert action. However, during the radiation dominated epoch, we find that $f(a)\propto a^{-4}$, while the dominant term of the scalar curvature is that of dust like matter, i.e. $R(a)\propto a^{-3}$. Therefore, during the radiation epoch, our solution behaves as $f(R)\propto R^{4/3}$. This model is stable until the present time but as $f_{RR}$ and $m_{eff}^2$ become negative in the future, see Fig.~\ref{derivs_darkmatter}, it will therefore suffer from the known Dolgov-Kawasaki instability \cite{Dolgov:2003px,Faraoni:2006sy}.

\begin{figure}[t]
	\includegraphics[width=\textwidth]{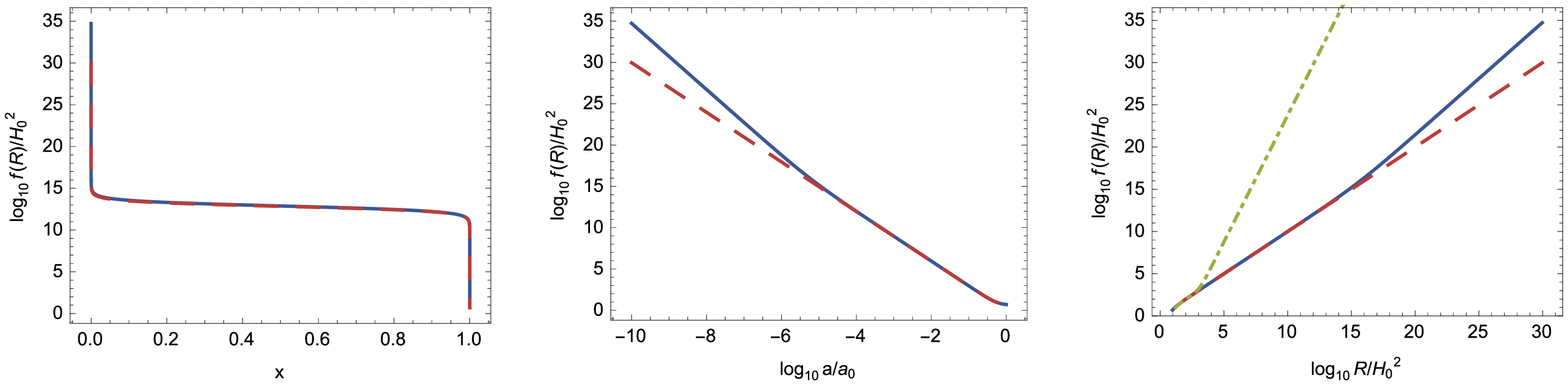}
	\caption{\label{numerical_darkradiation}The numerical solution of $f(R)$ (blue curve), that accounts for the matter content corresponding to the mGCG with $\beta=\frac13$ in the presence of a dark matter and baryonic matter and with a radiation content, versus the action $R-2\Lambda$ (red dashed curve). The numerical integration has been carried out imposing that at present $f(R_0)=R_0-2\Lambda$ and $f_R(R_0)=1$. The values of the parameters are given in Table.~\ref{back_para_tab} and Eqs.~\eqref{rad_tot}-\eqref{As_tot}. The green dot-dashed line in the right plot shows the third order expansion of $f(R)$ around $R_0$ obtained from the cosmographic approach, cf. Eq.~\eqref{cosmo_results}.
	}
\end{figure}

\begin{figure}[t]
	\includegraphics[width=\textwidth]{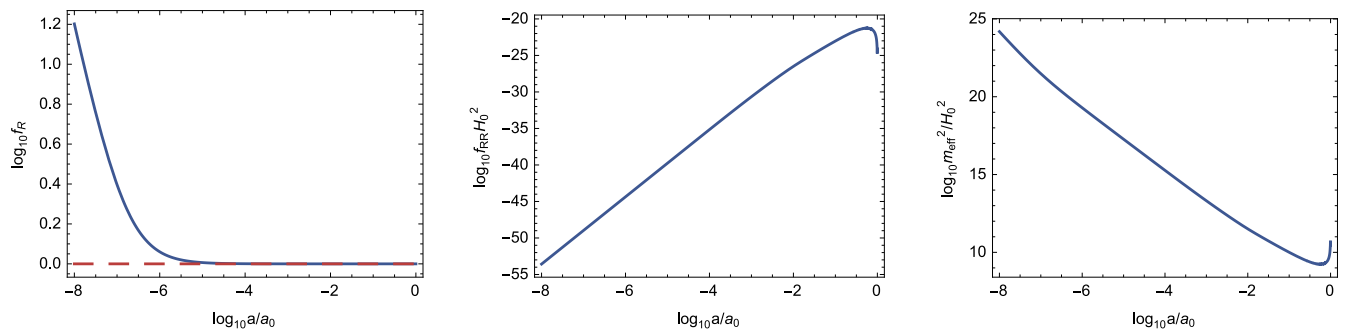}
	\caption{\label{derivs_darkradiation}The evolution of: $f_R$ (left panel); $f_{RR}$ (middle panel); and $m_{eff}^2$ (right panel); for the numerical solution in Fig.~\ref{numerical_darkradiation} (blue curves). A dashed red curve indicates the value of the quantity in GR. While $f_R$ remains positive throughout the entire evolution, $f_{RR}$ and $m_{eff}^2$ become negative in the future, when $a> a_0$, which will give rise to instabilities in the future.  The values of the parameters are given in Table.~\ref{back_para_tab} and Eqs.~\eqref{rad_tot}-\eqref{As_tot}.
	}
\end{figure}

We can as well address the possibility that $f(R)$ mimics only the dark radiation component and dark energy. In that case, the background evolves as
\begin{align}
	\label{background_rad}
	3H^2 = \rho_{r,0}\left(\frac{a_0}{a}\right)^{3} + \rho_{m,0}\left(\frac{a_0}{a}\right)^{3} + \rho_{ch,0}\left[A_s + (1-A_s)\left(\frac{a_0}{a}\right)^{4(1+\alpha)}\right]^{\frac{1}{1+\alpha}},
\end{align}
where $\rho_{r,0}$ is the present energy density of true radiation, i.e. photons and neutrinos. Considering that the three known species of neutrinos are still relativistic, the present day energy density of radiation should be \cite{Mangano:2005cc}
\begin{align}
	\Omega_{r,0} = \left[1 +\frac{7}{8}\left(\frac{4}{11}\right)^{4/3}N_{eff}^{(\nu)}\right]\Omega_{\gamma,0},
\end{align}
where $\Omega_\gamma$ is the relative energy density of photons and $N^{(\nu)}_{eff}=3.046$. Here, the small deviation from $N_{eff}=3$ comes from the effects of non-instantaneous neutrino decoupling from the photon-baryon plasma \cite{Mangano:2005cc}. Nevertheless, recent results have set the best fit value of $N_{eff}=3.15$ \cite{Ade:2015xua}. The difference $N_{eff}^{(dr)}\equiv N_{eff}-N^{(\nu)}_{eff}$ thus corresponds to a small excess of the radiation density, which is referred to as dark radiation \cite{Archidiacono:2011gq, Menestrina:2011mz, Ade:2015bva}. In order for the mGCG to account for this dark radiation component, we set
\begin{align}
	\label{rad_tot}
	&\Omega_{r,0}^{(tot)} = \frac{\Omega_{m,0}}{1+z_{eq}}.
	\\
	\label{rad_true}
	&\Omega_{r,0} = \left[1-\frac{\frac{7}{8}\left(\frac{4}{11}\right)^{4/3}N_{eff}^{(dr)}}{1+\frac{7}{8}\left(\frac{4}{11}\right)^{4/3}N_{eff}}\right]\Omega_{r,0}^{(tot)}.
	\\
	\label{rad_dark}
	&\Omega_{dr,0} = \frac{\frac{7}{8}\left(\frac{4}{11}\right)^{4/3}N_{eff}^{(dr)}}{1+\frac{7}{8}\left(\frac{4}{11}\right)^{4/3}N_{eff}}\Omega_{r,0}^{(tot)}.
	\\
	\label{chap_dark}
	&\Omega_{ch,0} = 1-\Omega_{r,0}-\Omega_{m,0}.
	\\
	\label{As_tot}
	&A_s = 1 - \left(\Omega_{dr,0}/\Omega_{ch,0}\right)^{1+\alpha}.
\end{align}
Following the previous procedure, we obtain the numerical solution for $f(R)$. The results obtained for the function $f(R)$, $f_R$, and $f_{RR}$, and $m_{eff}^2$ are plotted in Figs.~\ref{numerical_darkradiation} and \ref{derivs_darkradiation} . Again, this model is stable until the present time but as $f_{RR}$ and $m_{eff}^2$ become negative in the future, see Fig.~\ref{derivs_darkradiation}, it will therefore suffer from the known Dolgov-Kawasaki instability \cite{Dolgov:2003px,Faraoni:2006sy}.


In the next sections, we will constrain the two models introduced above. Our constraints will be based on a cosmographic approach and on a perturbative approach derived from the predicted matter power spectrum for each model.
%
%

\section{The Cosmographic approach}
\label{Sec4}

In this section, we review the cosmographic approach \cite{Capozziello:2008qc,Visser:2004bf,Cattoen:2007sk,Cattoen:2008th,BouhmadiLopez:2010pp,Trockel:2011sd,Capozziello:2011tj,Luongo:2011zz,Lavaux:2011yh,Xia:2011iv,Demianski:2012ra,Luongo2012,Aviles:2012ay,Bamba:2012cp,Neben:2012wc,Aviles:2012ir,Capozziello:2013wha,Luongo:2013rba,Gruber:2013wua,Aviles:2014rma,Aviles:2013zz} and apply it to the models introduced in the previous section. These models describe the evolution of the universe since the BBN and are of two types: (i) an $f(R)$ model with a dark matter and baryonic content and where the effective $f(R)$ energy density is described by a mGCG with $\beta=1/3$ and accounts for all the radiation content of the universe, i.e., we fix $\rho_{r,0}=0$ in Eq.~\eqref{background_rad}. (ii) a model similar to the one in (i) where the $f(R)$ accounts only for dark radiation, i.e. $\Omega_{r,0}$ is given by Eq.~\eqref{rad_true}.

One of the virtues of the observational cosmographic constraints is that, theoretically, they should be independent of the gravitational theory used at hand (for works on the cosmographic approach in $f(R)$ theories of gravity see for example \cite{Capozziello:2008qc,BouhmadiLopez:2010pp,Aviles:2012ir}). On the other hand, the cosmological expansion of the models we are studying can be conformally mapped to an $f(R)$ or a GR model (from a background point of view, and again within a theoretical framework). 
Therefore, these models, which correspond to different theories with the same cosmological expansion, are compatible within the cosmological approach iff there is a set of parameters in both models (modified gravity and GR) that gives the same cosmographic set of parameters.

One of the key issues we are tackling is the presence of radiation in the universe. We will therefore extend the cosmographic approach, within GR and $f(R)$, to include the contribution of radiation. Although we do not expect the current cosmographic parameters to be modified extensively by the inclusion of radiation, the latter has been included for consistency. 
It is important to stress that $f(R)$ (as soon as $f(R)\neq R$) and GR differ for the number of gravitational degrees of freedom. In order to extract the same cosmographic information, we need to know how this further degrees of freedom behave in the Einstein frame (i.e. how $f(R)$ gravity can be interpreted in term of GR plus scalar fields). This kind of analysis strictly depends on the gauge invariance, the weak field limit and other issues that have to be carefully considered. For a detailed discussion see \cite{Stabile:2013eha}. Here we discuss the cosmographic approach taking care of reproducing the main cosmographic parameters in term of $f(R)$ quantities (see Ref.~\cite{Capozziello:2008qc}). In this sense, we will remain strictly in the Jordan frame.

\subsection{Basic formulas of Cosmography}

We begin by expanding the scale factor of the FLRW metric as follows \cite{Capozziello:2008qc}
\begin{align}
	\label{a_expansion}
	\frac{a}{a_0} = 1 + H_0(t-t_0) - \frac{q_0}{2}H_0^2 (t-t_0)^2 + \frac{j_0}{3!}H_0^3(t-t_0)^3 + \frac{s_0}{4!}H_0^4(t-t_0)^4 + \frac{l_0}{5!}H_0^5(t-t_0)^5 + O\left((t-t_0)^6\right).
\end{align}
The cosmographic parameters introduced in the previous equation can be defined as
\begin{align}
	H &= \frac{1}{a}\frac{da}{dt},\\
	q &= -\frac{1}{a}\frac{d^2a}{dt^2}H^{-2}, \\
	j &= \frac{1}{a}\frac{d^3a}{dt^3}H^{-3}, \\
	s &= \frac{1}{a}\frac{d^4a}{dt^4}H^{-4}, \\
	l &= \frac{1}{a}\frac{d^5a}{dt^5}H^{-5}.
\end{align}
It is important to stress that we are not assuming any cosmological model but only the fact that the universe is homogeneous and isotropic.
Using the previous definitions we can write the time derivatives of the Hubble parameter as \cite{Capozziello:2008qc}
\begin{align}
	\label{cosm_Ht}
	\dot{H} &= -H^2(1+q), \\
	\label{cosm_Htt}
	\ddot{H} &= H^3(j + 3q +2), \\
	\label{cosm_Httt}
	\dddot{H} &= H^4\left[s - 4j - 3q(q+4) - 6\right], \\
	\label{cosm_Htttt}
	\ddddot{H} &= H^5\left[l - 5s + 10(q+2)j + 30(q+2)q + 24\right].
\end{align}
In appendix~\ref{app:cosmography}, we present the expression of the cosmographic parameters in a universe filled by dust-like matter, radiation, and a generic third fluid that accounts for dark energy. As an example, we apply these results to the $\Lambda$CDM model with a radiation component, where the equations of evolution for the universe are
\begin{align}
	3H^2 &= \rho_{m} + \rho_r + \rho_\Lambda, \\
	2\dot{H} &= -\rho_{m} - \frac{4}{3}\rho_r.
\end{align}
The expression of $q_0$, $j_0$, $s_0$, and $l_0$ as functions of $\Omega_{r,0}$ and $\Omega_{m,0}$ are
\begin{align}
	\label{cosmo_parameters}
	q_0 &= -1 +\frac{3}{2}\Omega_{m,0} + 2\Omega_{r,0}, \\
	j_0 &= 1 + 2\Omega_{r,0}, \\
	s_0 &= 1 - \frac{9}{2}\Omega_{m,0} - \left(12 + 3\Omega_{m,0}+4\Omega_{r,0}\right)\Omega_{r,0}, \\
	l_0 &= 1 + 3\Omega_{m,0} + \frac{27}{2}\Omega_{m,0}^2 + \left(28+72\Omega_{m,0} + 76\Omega_{r,0}\right)\Omega_{r,0} .
\end{align}
These reduced to the expressions in \cite{Xia:2011iv,Capozziello:2008qc,Aviles:2012ir} when we set $\Omega_{r,0}=0$.
Let us now deduce the expressions for the cosmographic parameters in a model where the background is given by Eq.~\eqref{background_rad}, i.e.,
\begin{align}
	3H^2 &= \rho_m + \rho_r+ \rho_{ch}, \\
	2\dot{H} &= -\rho_m - \frac{4}{3}\rho_r - \frac{4}{3}\rho_{ch}\left[ 1 - A_s\left(\frac{\rho_{ch,0}}{\rho_{ch}} \right)^{1+\alpha}\right].
\end{align}
In the second equation we have used Eq.~\eqref{mGCG_eos} and the definition of $A_s$ to write $p_{ch}$ in terms of $\rho_{ch}$ and the mGCG parameters $\alpha$ and $A_s$.
The cosmographic parameters now read
\begin{align}
	\label{q0_mGCG}
	q_0 =& -1 + \frac{3}{2}\Omega_{m,0} + 2\Omega_{r,0}+ 2(1-A_s) \left(1-\Omega_{m,0}-\Omega_{r,0}\right),
	\\
	\label{j0_mGCG}
	j_0 =& 1 + 2\Omega_{r,0}
		+ 2(1-A_s)\left(1+4\alpha A_s\right)\left(1-\Omega_{m,0}-\Omega_{r,0}\right),
	\\
	\label{s0_mGCG}
	s_0 =& 1 - \frac{9}{2}\Omega_{m,0} - \left(12+3\Omega_{m,0}+4\Omega_{r,0}\right)\Omega_{r,0}
	\nonumber\\&
	-(1-A_s)
	\left\{
		4\left[3 - 8\alpha^2A_s + 8\alpha(1+\alpha)A_s^2\right]
		+3(1+4\alpha A_s)\Omega_{m,0}
		+8(1+2\alpha A_s)\Omega_{r,0}
	\right\}\left(1-\Omega_{m,0}-\Omega_{r,0}\right)
	\nonumber\\&
	-4(1-A_s)^2(1+4\alpha A_s)\left(1-\Omega_{m,0}-\Omega_{r,0}\right)^2,
	\\
	\label{l0_mGCG}
	l_0 =& 1 + 3\Omega_{m,0} + \frac{27}{2}\Omega_{m,0}^2 + \left(28+72\Omega_{m,0} + 76\Omega_{r,0}\right)\Omega_{r,0}
	\nonumber\\&
	+4(1-A_s)
	\left\{
		7
		+4\alpha\left( 3 + 2\alpha + 8\alpha^2\right)A_s
		- 8\alpha\left(5+22\alpha+24\alpha^2\right)A_s^2
		+32\alpha\left(2+7\alpha+6\alpha^2\right)A_s^3
	\right.
	\nonumber\\&~~~~~~~~~~~~~~~~
	\left.
		+9
		\left[
			2
			+ \alpha(1-4\alpha)A_s
			+4\alpha(1+2\alpha)A_s^2
		\right]\Omega_{m,0}
	\right.
	\nonumber\\&~~~~~~~~~~~~~~~~
	\left.
		+2
		\left[
			19
			+2\alpha(5-12\alpha)A_s
			+24\alpha(1+2\alpha)A_s^2
		\right]\Omega_{r,0}
	\right\}\left(1-\Omega_{m,0}-\Omega_{r,0}\right)
	\nonumber\\&
	+4(1-A_s)^2
	\left[
		19 + 4\alpha(5-12\alpha)A_s
		+16\alpha(3+7\alpha)A_s^2
	\right]\left(1-\Omega_{m,0}-\Omega_{r,0}\right)^2.
\end{align}
Notice that for $A_s=1$, the mGCG behaves as a cosmological constant and we recover the results of $\Lambda$CDM. For $A_s=0$, the mGCG behaves as pure radiation. Therefore, we recover the results in Eq.~\eqref{cosmo_parameters}, where the total energy density of radiation is $\Omega^{tot}_{r,0}=\Omega_{r,0}+\left(1-\Omega_{m,0}-\Omega_{r,0}\right)=1-\Omega_{m,0}$. These formulas apply to both of the models introduced in Sec.~\ref{Sec3}, where we set $\Omega_{r,0}=0$ for the model where the mGCG accounts for all the radiation content of the universe. 

\subsection{Cosmography within $f(R)$ gravity}

The cosmographic approach can be straightforwardly adapted to $f(R)$ gravity as soon as the cosmographic parameters are recast in terms of $f(R)$ derivatives. We begin by writing the scalar curvature and its time derivatives in terms of the cosmographic parameters \cite{Capozziello:2008qc,BouhmadiLopez:2010pp, Aviles:2012ir}. We obtain
\begin{align}
	\label{cosm_R}
	R &= 6H^2(1-q),\\
	\label{cosm_Rt}
	\dot{R} &= 6H^3(j-q-2),\\
	\label{cosm_Rtt}
	\ddot{R} &= 6H^4(s + q^2 + 8q +6),\\
	\label{cosm_Rttt}
	\dddot{R} &= 6H^5\left[l - s - 2(q+4)j - 6(3q+8)q - 24\right].
\end{align}

Using the modified Friedmannn and Raychaudhury equations, we obtain expressions for $f$ and $f_{RRR}$ in terms of the matter energy density, $\rho$, and the derivatives $f_R$ and $f_{RR}$ \cite{Capozziello:2008qc,BouhmadiLopez:2010pp}
\begin{align}
	\label{cosm_f}
	f &= 2\rho + (R-6H^2)f_{R} - 6H\dot{R}f_{RR},
	\\
	\label{cosm_f'''}
	f_{RRR} &= -\frac{(1+w)\rho+2\dot{H}f_{R}+\left(\ddot{R}-H\dot{R}\right)f_{RR}}{(\dot{R})^2}.
\end{align}
Here, $\rho$ will account for the total matter content of the universe. For the first model introduced in Sec.~\ref{Sec3B}, this corresponds to dark and baryonic matter, while in the second model introduced in Sec.~\ref{Sec3B}, we consider dark and baryonic matter plus radiation. The parameter $w$ is defined in the standard way $w=p/\rho$, i.e., the quotient between the total matter pressure and energy density.

In order to obtain a third equation, we differentiate the Raychaudhuri equation once with respect to the cosmic time and obtain
\begin{align}
	\label{Raych_deriv}
	\ddot{H} =&
		\frac{1+w}{2}\left[3\left(1+c_s^2\right)H + \dot{R}\frac{f_{RR}}{f_{R}}\right]\frac{\rho}{f_{R}}
		- \frac{1}{2}\left(\dddot{R} - H\ddot{R} - \dot{H}\dot{R}\right)\frac{f_{RR}}{f_{R}}
		+ \frac{1}{2}\left[\ddot{R}\dot{R}-H(\dot{R})^2\right]\left(\frac{f_{RR}}{f_{R}}\right)^2
	\nonumber\\&
	- \frac{1}{2}\left[3\ddot{R}\dot{R}-H(\dot{R})^2\right]\frac{f_{RRR}}{f_{R}}
	+ \frac{(\dot{R})^3}{2}\frac{f_{RRR}f_{RR}}{(f_{R})^2} 
	- \frac{(\dot{R})^3}{2}\frac{f_{RRRR}}{f_{R}}.
\end{align}
The squared speed of sound is defined as $c_s^2\equiv dp/d\rho$. Since at the present time the deviation from GR should be small, we can safely approximate the function $f(R)$ by its Taylor expansion around $R_0$ up to third order, i.e.
\begin{align}
	f(R) \approx f(R_0) + f_{R}(R_0)(R-R_0) + \frac{f_{RR}(R_0)}{2}(R-R_0)^2 + \frac{f_{RRR}(R_0)}{6}(R-R_0)^3 + \mathcal{O}(R-R_0)^4.
\end{align}
With this assumption in mind, we can ignore the term in the fourth derivative $f_{RRRR}$ in Eq.~\eqref{Raych_deriv}. After using Eq.~\eqref{cosm_f'''} to eliminate $f_{RRR}$ in Eq.~\eqref{Raych_deriv}, and sorting the terms by equal power in $f_{RR}$, we can rewrite Eq.~\eqref{Raych_deriv} as
\begin{align}
	&\left[
		\left(3\ddot{R} - H\dot{R}\right)\left(\ddot{R}-H\dot{R}\right)
		- \left(\dddot{R}\dot{R} + \dot{H}(\dot{R})^2 - H\dot{R}\ddot{R}\right)
	\right]\frac{f_{RR}}{f_{R}}
	\nonumber\\&~~~~~~
	+ \left[3\ddot{R} + \left(1+w\right)\left(2+3c_s^2\right)H\dot{R}\right]\frac{\rho}{f_{R}}
	+ 6\ddot{R}\dot{H}
	- 2H\dot{H}\dot{R}
	- 2\ddot{H}\dot{R}
	=0.
\end{align}
Thus we now have a linear equation in $f_{RR}$, which, after a simple manipulation gives \cite{Capozziello:2008qc}
\begin{align}
	\label{cosm_f''}
	f_{RR} = \frac{\left[3\ddot{R} + \left(1+w\right)\left(2+3c_s^2\right)H\dot{R}\right]\rho
	+ \left(
		6\ddot{R}\dot{H}
		- 2H\dot{H}\dot{R}
		- 2\ddot{H}\dot{R}
	\right)f_{R}}{\dddot{R}\dot{R} + \dot{H}(\dot{R})^2 - H\dot{R}\ddot{R} - \left(3\ddot{R} - H\dot{R}\right)\left(\ddot{R}-H\dot{R}\right)}.
\end{align}
We can express the current value of $f$, $f_{RR}$ and $f_{RRR}$ in terms of the cosmographic parameters by substituting Eqs.~\eqref{cosm_Ht}-\eqref{cosm_Htttt} and \eqref{cosm_R}-\eqref{cosm_Rttt} in Eqs.~\eqref{cosm_f}, \eqref{cosm_f'''} and \eqref{cosm_f''}. If we consider that the universe is filled by dust like matter, $\rho_m$, and radiation, $\rho_r$, then the expressions obtained are
\begin{align}
	\label{f0}
	\frac{f(R_0)}{6H_0^2} &= -\frac{\mathcal{A}_0\Omega_m+\mathcal{B}_0 + \mathcal{C}_0\Omega_r}{\mathcal{D}}, \\
	\label{f2}
	\frac{f_{RR}(R_0)}{(6H_0^{2})^{-1}} &= -\frac{\mathcal{A}_2\Omega_m+\mathcal{B}_2+ \mathcal{C}_2\Omega_r}{\mathcal{D}}, \\
	\label{f3}
	\frac{f_{RRR}(R_0)}{(6H_0^2)^{-2}} &= -\frac{\mathcal{A}_3\Omega_m+\mathcal{B}_3+ \mathcal{C}_3\Omega_r}{(j_0-q_0-2)\mathcal{D}},
\end{align}
where the coefficients $\mathcal{A}_i$, $\mathcal{B}_i$ and $\mathcal{D}$ are defined as
\begin{align}
	\mathcal{A}_0 =& \left(j_0-q_0-2\right)l_0
		- \left[3s_0 + 7j_0 + 6q_0^2 + 41q_0 + 22\right]s_0 \nonumber\\
		&- \left[(3q_0+16)j_0 + 20q_0^2 + 64q_0 + 12\right]j_0
		-3q_0^4
		-25q_0^3
		-96q_0^2
		-72q_0
		-20, \\
	\mathcal{B}_0 =& -\left(q_0j_0-q_0^2-2q_0\right)l_0
		+ \left[3q_0s_0 + (4q_0+6)j_0 +6q_0^3 + 44q_0^2 + 22q_0 - 12\right]s_0 \nonumber\\
		&+ \left[ 2j_0^2 + (3q_0^2 + 10q_0 -6)j_0 + 17q_0^3 +52q_0^2 +54q_0 + 36\right]j_0
		+ 3 q_0^5 + 28q_0^4 \nonumber\\
		&+ 118q_0^3 + 72q_0^2 - 76q_0 - 64,
	\\
	\mathcal{C}_0 =&\left(j_0-q_0-2\right)l_0
		-\left[3s_0 + 10j_0 + 6q_0^2+ 38 q_0 + 16\right]s_0
		\nonumber\\&
		- \left[\left(3q_0+22\right)j_0 + 23q_0^2 + 76q_0 + 6\right]j_0
		- 3q_0^4 - 22q_0^3 - 72q_0^2 - 30q_0 - 8,
	\\
	\mathcal{A}_2 =& 3\left[3s_0 + 2j_0 + 3q_0^2 + 22q_0 +14\right], 
	\\
	\mathcal{B}_2 =& -2\left[3(1+q_0)s_0 + (j_0+q_0-1)j_0 + 3q_0^3 + 25q_0^2 +37q_0 +16\right], 
	\\
	\mathcal{C}_2 =& 12\left[s_0+j_0+q_0^2 + 7q_0 +4\right], 
	\\
	\mathcal{A}_3 =& -3\left[l_0 + s_0 - (3q_0+12)j_0 - 15q_0^2 - 26q_0 - 4\right],
	\\
	\mathcal{B}_3 =& 2\left[(1+q_0)l_0 + (j_0 + q_0)s_0 - (j_0 + 2q_0^2 + 6q_0 + 3)j_0 - 15q_0^3 - 42q_0^2 - 39q_0 - 12\right], 
	\\
	\mathcal{C}_3 =& - 4\left[l_0 + 2s_0 + \left(3q_0+ 13\right)j_0 + 14q_0^2 + 17q_0 - 4\right], 
	\\
	\label{cosm_D}
	\mathcal{D} =& -\left(j_0-q_0-2\right)l_0
		+ \left[3s_0 - 2j_0 + 6q_0^2 + 50q_0 + 40\right]s_0\nonumber\\
		&+\left[(3q_0+10)j_0 + 11q_0^2 + 4q_0 -18\right]j_0
		+ 3q_0^4
		+ 34q_0^3
		+ 180q_0^2
		+ 246q_0
		+ 104.
\end{align}
Notice that here we assume that $f_R(R_0)=1$ and use Eq.~\eqref{cosm_f''} to eliminate $f_{RR}$ in Eqs.~\eqref{cosm_f} and \eqref{cosm_f'''}. In addition, the $0$ subscript in the coefficients $\mathcal A$, $\mathcal B$, and $\mathcal C$, stands not for evaluation at the present time, but to the order of the $f(R)$ derivative to which they correspond (c.f. Eqs.~\eqref{f0}-\eqref{f3}). The expression obtained for $\mathcal A_i$ and $\mathcal B_i$ are the same as the ones obtained in \cite{Capozziello:2008qc,BouhmadiLopez:2010pp}.

Let us say a few more words on how we can constrain our model. There are several options:
\begin{enumerate}
\item We fit the two models introduced on the previous subsection using for example data from SNeIa, BAO and CMB. Once the cosmological parameters are fitted, we can obtain the corresponding cosmographic parameters given in Eqs.~\eqref{q0_mGCG}-\eqref{l0_mGCG}. Then by using the expressions \eqref{f0}-\eqref{cosm_D} we fully determine our model at present and on the near past.

\item We get the cosmographic parameters by using data for example from SNeIa, BAO and CMB. Then by inverting the relations \eqref{q0_mGCG}-\eqref{l0_mGCG}, we would obtain $\Omega_{m,0}$, $A_s$ and $\alpha$.
Finally, by plugging those values and the cosmographic parameters in Eqs. \eqref{f0}-\eqref{cosm_D}, the $f(R)$ model would be fully determined.

\item To get an order of magnitude of the parameters of our model, we can assume that at present our model mimics pretty much $\Lambda$CDM, therefore, $\Omega_{m,0}\sim 0.3065$; i.e. cf. latest Planck results \cite{Ade:2015xua}, $A_s\sim 1$ and $\alpha \sim 0$, therefore, $q_0$, $j_0$, $s_0$, and $l_0$ can be determined using Eqs.~\eqref{q0_mGCG}-\eqref{l0_mGCG} and again by using the expressions \eqref{f0}-\eqref{cosm_D}, we fully determine our model at present and on the near past.
\end{enumerate}

In the previous three approaches we can assume $\Omega_{r,0}^{(tot)}$, $\Omega_{r,0}$ and $\Omega_{dr,0}$ to be fixed, see eqs.~\eqref{rad_tot}, \eqref{rad_true}, and \eqref{rad_dark}, by the values of $\Omega_{m,0}$, $z_{eq}$, and $N_{eff}$ from the Planck Collaboration 2015 \cite{Ade:2015xua}. For our purpose, we use the third approach while considering the values of cosmographic parameters deduced from the data reported by the Planck Collaboration 2015 \cite{Ade:2015xua}, which can be considered as the best cosmological data available up to now.
We then obtain
\begin{align}
	q_0 &= - 0.54007,
	\quad&
	j_0 &= 1.0002,
	\quad&
	s_0 &= -0.38042,
	\quad&
	l_0 &= 3.1922,
\end{align}
and
\begin{align}
	\label{cosmo_results}
	\frac{f(R_0)}{6H_0^2} &= 0.57451,
	\quad&
	\frac{f_{RR}(R_0)}{(6H_0^{2})^{-1}} &= 1.4962\times10^{-16}, 
	\quad&
	\frac{f_{RRR}(R_0)}{(6H_0^2)^{-2}} &= 1.3017\times10^{-4}.
\end{align}
The values here obtained via the cosmographic approach are consistent with the numerical solution for $f(R)$ obtained in the previous section in the presence of dust and radiation, with the third order Taylor expansion of $f(R)$ around $R_0$ providing a very good approximation for the numerical solution at present, cf. Fig.~\ref{numerical_darkradiation}. Our further considerations will be based on this assumption.

%
%

\section{The Clumpy Universe}
\label{Sec5}
We next analyze the evolution of the scalar perturbations for our models, since the radiation epoch until the present time, and compare it with the evolution of perturbations in the concordance model, i.e., the $\Lambda$CDM model. In particular, we will look for the effects of the $f(R)$ modifications in the matter power spectrum as measured today. 

While cosmography is one of the many approaches to test the background/smooth universe, the matter power spectrum is a tool to test the clumpy universe, which we will use to analyze the models introduced in Sec.~\ref{Sec3}. The matter power spectrum provides information about the process of clustering of matter in the universe at all scales, i.e., from those corresponding to the early radiation dominated epoch/late reheating, to those currently exiting the horizon, which have recently reentered the horizon. In addition, it is a magnificent tool to break the possible degeneracy between GR and some modified theories of gravity at the background level. In summary, the matter power spectrum is a powerful tool to test cosmological models, and modified theories of gravity in particular. In order to obtain the matter power spectrum we need the evolution of the scalar perturbations, which we next start reviewing within an $f(R)$ setup. Coupling scalar perturbations with cosmography can be, in principle, an efficient tool to trace back the cosmic history.

\subsection{Scalar Perturbations}

The scalar part of the perturbed FLRW metric can be written in a gauge invariant way (which coincides with the Newtonian gauge) as
\cite{Mukhanov:1990me,Malik:2001rm,Dodelson:2003,Malik:2008im,Amendola:2010,Kurki-Suonio2012}
\begin{align}
	ds^2 = a^2\left[-\left(1+2\Phi\right)d\eta^2 + \left(1-2\Psi\right)\delta_{ij}dx^idx^j\right],
\end{align}
where $\Phi\left(\eta,x^i\right)$ and $\Psi\left(\eta,x^i\right)$ are the gauge invariant Bardeen potentials \cite{Bardeen:1980kt}. At first order in perturbations, we can write the energy-momentum tensor as
\cite{Mukhanov:1990me,Malik:2001rm,Dodelson:2003,Malik:2008im,Amendola:2010,Kurki-Suonio2012}
\begin{align}
	\label{T_pert}
	T^0_0 &= -(\rho+\delta\rho),
	\nonumber\\
	T^0_i &= -(\rho+p)v_i,
	\nonumber\\
	T^i_j &=(p+\delta p)\delta^i_j,
\end{align}
where $\delta\rho$ is the energy density perturbation, $\delta p$ is the pressure perturbation and $v$ is the peculiar velocity. Here, we do not take into account the effects of the anisotropic stress.

In a metric $f(R)$ theory, the $(ij)$ component of the perturbed Einstein equations relates the potentials $\Phi$ and $\Psi$ with the perturbation of $f_R$ \cite{Song:2006ej,Carroll:2006jn,Pogosian:2007sw,delaCruzDombriz:2008cp}
 \begin{align}
 	\left(\Psi -\Phi\right)f_R = \delta f_R = f_{RR}\delta R.
 \end{align}
Thus the equality between the two potentials that we see in GR is in general no longer true in an $f(R)$ theory and just as $f_R$ can be interpreted as a new degree of freedom in the background evolution, so does $\delta f_R$, which acts as a new degree of freedom at the perturbative level. At this point we introduce the new set of variables $\Psi^+$ and $\Xi$,
\begin{align}
	\Psi^+ \equiv \frac{\Phi+\Psi}{2},
	~~~~~~~~~~~~~~~~
	\Xi \equiv \frac{\delta f_R}{f_R} = \Psi-\Phi.
\end{align}
Notice that $\Psi^+$ corresponds to the variable $\Phi^+$ in Ref.~\cite{Pogosian:2007sw}, while $\Xi$ corresponds to the variable $\chi$ of the same reference divided by $f_R$. We choose the variable $\Xi$ instead of $\chi$, because this choice directly reflects the difference between the two potentials and allows for a better control of the numerical integrations, even when the theory could deviate considerably from GR ($f_R\gg 1$). Using these variables, we can write the $(00)$ and $(i0)$ components of the perturbed Einstein equations as \cite{delaCruzDombriz:2008cp,Carroll:2006jn,Pogosian:2007sw}
\begin{align}
	\label{EinsPert00}
	\left[1+\frac{1}{2}\frac{\left(f_R\right)_N}{f_R}\right]\left(\Psi^+ \right)_N
	+\left[1+\frac{\left(f_R\right)_N}{f_R} + \frac{k^2}{3\mathcal{H}^2}\right]\Psi^+ 
	+\frac{1}{4}\frac{\left(f_R\right)_N}{f_R}\left(\Xi\right)_N
	-\frac{1}{2}\left[1 - \mathcal{H}_N +2\frac{\left(f_R\right)_N}{f_R}\right]\Xi
	&= - \frac{a^2\kappa^2}{6\mathcal{H}^2f_R}\delta\rho,
\end{align}
\begin{align}
	\left(\Psi^+\right)_N 
	+ \left[1+\frac{1}{2}\frac{\left(f_R\right)_N}{f_R}\right]\Psi^+
	-\frac{3}{4}\frac{\left(f_R\right)_N}{f_R}\Xi
	&= - \frac{a^2\kappa^2\left(\rho+p\right)}{2f_R}\frac{v}{\mathcal{H}}.
\end{align}
Here, an $N$-subscript indicates a derivative with respect to $N\equiv\log(a)$. The previous two equations can be combined to obtain the evolution equations for $\Psi^+ $ and $\Xi$
\begin{align}
	\label{Psi+_eq}
	\left(\Psi^+ \right)_N &= -\Psi^+ - \frac{1}{4}\frac{\left(f_R\right)_N}{f_R}\left(2\Psi^+ - 3\Xi\right) - \frac{a^2\kappa^2(\rho+p)}{2f_R}\frac{v}{\mathcal H},
	\\
	\label{Xi_eq}
	\left(\Xi\right) _N &= \Xi + \frac{a^2\kappa^2(\rho+p)}{f_R}\frac{v}{\mathcal H} + \frac{1}{2}\frac{\left(f_R\right)_N}{f_R}\left(2\Psi^+ - 3\Xi\right) - \frac{2}{3\mathcal H^2}\frac{f_R}{\left(f_R\right)_N}\left[2k^2\Psi^+ + \frac{a^2\kappa^2}{f_R}\left(\delta\rho-3\mathcal{H}(\rho+p)v\right)\right]
	\nonumber\\
	&+\frac{2}{\mathcal{H}}\frac{f_R}{\left(f_R\right)_N}\left(\mathcal H - \mathcal H_N\right)\Xi.
\end{align}
Eqs. \eqref{Psi+_eq} and \eqref{Xi_eq} are equivalent to the ones obtained in Ref.~\cite{Pogosian:2007sw}. Notice, however, that we have used the evolution equation for $\Psi^+$ \eqref{Psi+_eq} to eliminate the dependence of $(\Xi)_N$ on $(\Psi^+)_N$ in Eq.~\eqref{Xi_eq}.

Having obtained the evolution equation for the metric perturbations, we now require differential equations that dictate the evolution of the perturbed matter quantities, namely $\delta\rho$ and $v$. For a collection of $I$ perfect fluids, each with an energy-momentum tensor of the type \eqref{T_pert}, we can define the total, $\delta$, and individual, $\delta^{(i)}$, relative density perturbation, respectively, as
\begin{align}
	\delta = \sum_{i=1}^I\frac{\rho^{(i)}}{\rho}\delta^{(i)},
	~~~~~~~~~~~~~~~~
	\delta^{(i)} \equiv \frac{\delta\rho^{(i)}}{\rho^{(i)}},
\end{align}
while the total and individual peculiar velocities are related by
\begin{align}
	v &= \sum_{i=1}^I \frac{\rho^{(i)}+p^{(i)}}{\rho+p}v^{(i)}.
\end{align}
By perturbing the conservation equations of the energy momentum tensor, we find that, for adiabatic and non-interacting fluids, each pair $\delta^{(i)}$ and $v^{(i)}$ satisfies 
\cite{Mukhanov:1990me,Malik:2001rm,Dodelson:2003,Malik:2008im,Amendola:2010,Kurki-Suonio2012}
\begin{align}
	\label{deltai_eq}
	&\left(\delta^{(i)}\right)_N + 3\left(c_s^{(i)2}-w^{(i)}\right) - (1+w^{(i)})k^2\frac{v^{(i)}}{\mathcal H}= 3(1+w^{(i)})\left(\Psi^+ + \frac{1}{2}\Xi\right)_N,
	\\
	\label{vi_eq}
	&\left(v^{(i)}\right)_N + \left(1-3c_s^{(i)2} \right)v^{(i)} + \frac{c_s^{(i)2}}{1+w^{(i)}}\frac{\delta^{(i)}}{\mathcal H} = - \frac{1}{\mathcal H}\left(\Psi^+ -\frac{1}{2} \Xi\right).
\end{align}
Here, $w^{(i)}\equiv p^{(i)}/\rho^{(i)}$ and $c_s^{(i)2}\equiv p^{(i)}_N/\rho^{(i)}_N$ are, respectively, the state parameter and the squared speed of sound of the $(i)$-fluid. 

Eqs. \eqref{Psi+_eq}, \eqref{Xi_eq}, \eqref{deltai_eq}, and \eqref{vi_eq} form a closed set of equations that allow us to evolve the perturbation variables since the early radiation epoch until the present time. Once the present day value of the perturbation variables is computed, they can be related with observable quantities.

\subsection{Matter Power Spectrum}
The perturbative analysis carried on the previous subsection will allows us to see how the $f(R)$ corrections affect the matter power spectrum of dark matter $\mathcal{P}_{\delta_{m}}$ \cite{Liddle:1993fq,Liddle:2000}.
Notice however that the correct definition of $\mathcal{P}_{\delta_{m}}$ uses the energy density perturbation $\delta_m$ in the comoving gauge (for a discussion of this topic see e.g. Refs.~\cite{Wands:2009ex,Bruni:2011ta}), while the variables used here correspond to the Newtonian gauge. In order to account for this gauge difference, we calculate the matter power spectrum as
\begin{align}
	\mathcal{P}_{\delta_{m}} \equiv \langle|\delta_{m}^{(com)}|^2\rangle = \langle|\delta_{m} - 3\mathcal{H}v_{m}|^2\rangle.
 \end{align} 

\begin{figure}[t]
	\includegraphics[width=\textwidth]{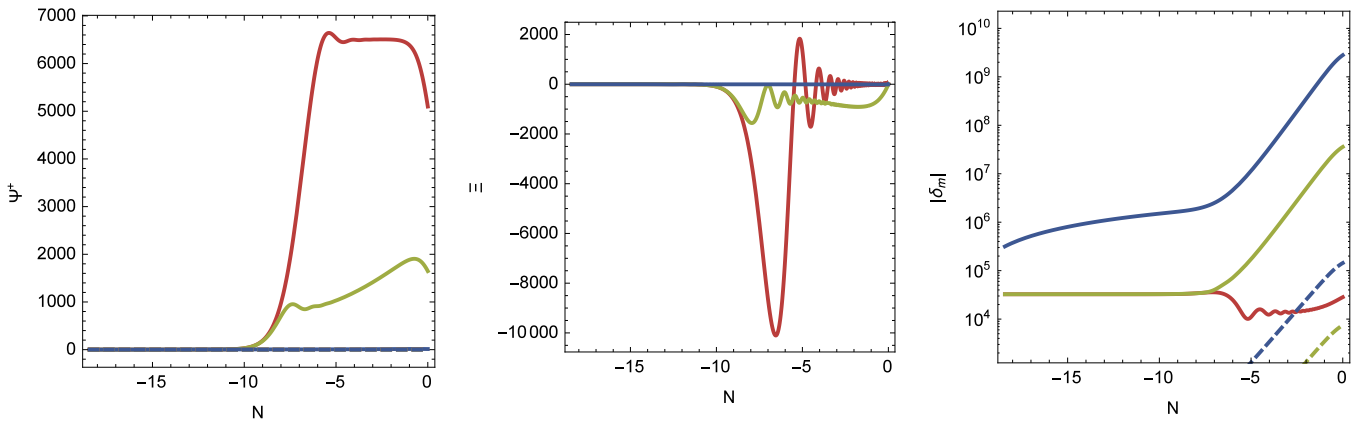}
	\caption{\label{Mat_Pert_Evol}The evolution of the perturbation variables: $\Psi^+$ (left); $\Xi$ (middle); $\delta_m$ (right); since the radiation dominated epoch until the present time for different values of $k$: $k=2\times10^{-4}$ Mpc$^{-1}$ (red); $k=2\times10^{-2}$ Mpc$^{-1}$ (green); $k=2$ Mpc$^{-1}$ (blue). The evolution of the same variables in the $\Lambda$CDM model is plotted in dashed lines. On the left panel, the value of $\Psi^+$ in the $\Lambda$CDM model is much smaller than in the $f(R)$ model, as such the dashed lines appear almost superimposed with $\Psi^+=0$. Since the perturbation $\Xi$ vanishes in GR, the dashed lines do not appear on the middle panel. On the right panel, the value of the matter perturbation in GR is several orders of magnitude smaller that the one in the $f(R)$ model, as such in GR only the modes with higher comoving wave number grow enough to appear on the plot. The $f(R)$ effective energy density accounts for all the radiation content of the universe (the first model introduced in Sec.~\ref{Sec3B}).}
\end{figure}

We will first analyze a model corresponding to a universe filled with dust-like matter and where the effective energy density of $f(R)$ mimics a mGCG, whose energy density is given in Eq.~\eqref{mGCG_of_a} with $\beta=1/3$. Therefore, the $f(R)$ model accounts for all the radiation content of the universe and drives its late time acceleration.
In Fig.~\ref{Mat_Pert_Evol}, we present the results obtained for the evolution of the perturbation variables $\Psi^+$, $\Xi$, and $\delta_m$, for different wave-numbers. The initial conditions
\begin{align}
	\label{pert_initcond}
	\Psi^+(N_{ini})=1,
	~~~~~~
	\left(\Psi^+\right)_N(N_{ini})=0,
	~~~~~~
	\Xi(N_{ini})=0,
	~~~~~~
	\left(\Xi\right)_N(N_{ini})=0,
\end{align}
were chosen so as to match the initial conditions in the $\Lambda$CDM model with a radiation component with standard single field inflationary conditions, i.e., the field $\Psi^+$ is initially constant and the modes are well outside of the horizon. During the radiation dominated epoch, the $f(R)$ corrections increase the value of the metric perturbations by several orders of magnitude, in stark contrast with what happens in GR. Furthermore, from Eqs.~\eqref{EinsPert00} and \eqref{pert_initcond}, we find that the initial value of the density perturbation $\delta$ is related to $\Psi^+(N_{ini})$ as
\begin{align}
	\label{init_delta_val}
	\delta_m(N_{ini}) = 
	- \frac{6\mathcal{H}^2f_R}{a^2\kappa^2\rho}\left[1+\frac{\left(f_R\right)_N}{f_R} + \frac{k^2}{3\mathcal{H}^2}\right]\Psi^+ (N_{ini})
	.
\end{align}
In our model, we find that initially $(f_R)_N/f_R\ll1$, furthermore, given that initially all the relevant modes are outside of the horizon, we have that $k^2/(3\mathcal{H}^2)\ll1$. Therefore, the relation between $\delta$ and $\Psi^+$ is dictated by the factor $6\mathcal{H}^2f_R/a^2\kappa^2\rho$. In GR, the Friedmann equation tells us that $3\mathcal{H}^2=a^2\kappa^2\rho$ ($f_R=1)$ and the matter and metric perturbations have the same order of magnitude. However, in $f(R)$ gravity, if the matter energy density is initially sub-dominant with regards to the $f(R)$ effective energy density, the factor on the right-hand-side (rhs) of Eq.~\eqref{init_delta_val} can become large, and so does $\delta$. Before proceeding further, we would like to highlight the following. Applying the GR limit to Eq.~\eqref{init_delta_val} gives $\delta_m=-2\Psi^+$. In fact, this relation refers to the total matter energy density perturbation $\delta$. For a GR model, aside from matter, we have as well radiation, therefore we need two initial conditions for $\delta_m$ and $\delta_r$. Those are obtained by imposing the adiabaticity of the initial conditions, i.e. \cite{Dodelson:2003,Amendola:2010}
\begin{align}
	\frac{\delta_m}{1+w_m} = \frac{\delta_r}{1+w_r} = \frac{\delta}{1+w},
\end{align}
where $w_m=0$, $w_r=1/3$, and initially $w\approx1/3$. Therefore, in a GR model with matter and radiation and initial adiabatic perturbations we have the initial relation between the matter and the metric perturbations
\begin{align}
	\delta_m = \frac{3}{4}\delta_r=\frac{3}{4}\delta = -\frac{3}{2}\Psi^+.
\end{align}
The presence of radiation and the adiabaticity of the initial perturbations thus corrects the GR limit of Eq.~\eqref{init_delta_val} by a factor of $3/4$.

\begin{figure}[t]
	\includegraphics[width=\textwidth]{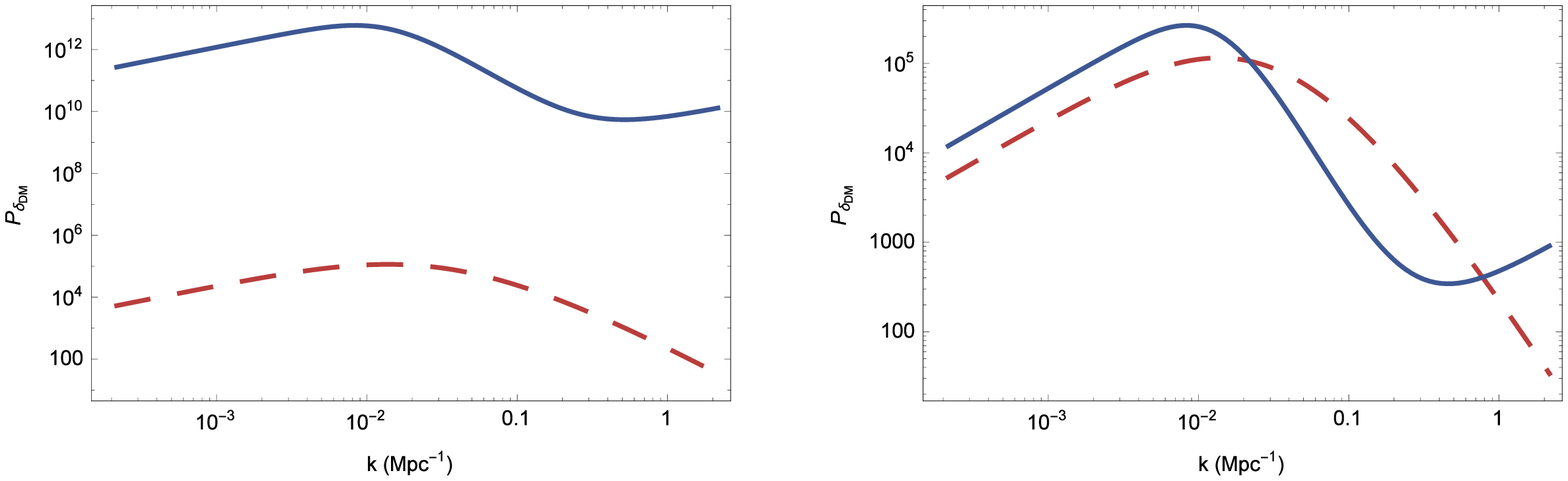}
	\caption{\label{Mat_spectra}The matter power spectrum $\mathcal P_{\delta_m}$ for the $f(R)$ model that accounts for all the radiation content of the universe (blue) and in the $\Lambda$CDM model (red discontinuous). On the left the matter power spectrum obtained for the initial conditions on Eq.~\eqref{pert_initcond}. On the right the matter power spectrum obtained by fine-tuning the initial conditions in order to minimize the deviations from the result of the $\Lambda$CDM model.}
\end{figure}

In our model the initial value of $\delta_m$ is several orders of magnitude higher than the one in GR, as can be seen on the plot on the right in Fig.~\ref{Mat_Pert_Evol}. As a result, the matter power spectrum obtained is very different from the one of the $\Lambda$CDM model, as is shown on the left plot of Fig.~\ref{Mat_spectra}. Besides the overall increase in the amplitude of the spectrum, we find that the $f(R)$ effects change the shape of the spectrum on the high $k$ regime of the spectrum ($k\geq k_{eq}\approx 0.01$~Mpc$^{-1}$). This is consist with the scale-dependent effects of $f(R)$, whose effects on the matter perturbations are stronger for higher $k$ \cite{Pogosian:2007sw,delaCruzDombriz:2008cp}. On the panel on the rhs of Fig.~\ref{Mat_spectra}, we show the matter power spectrum obtained by fine-tuning the initial conditions in order to minimize the deviations from the spectrum of the $\Lambda$CDM model. While we are able to obtain a relatively good fit on the low $k$ end of the spectrum, the same was not possible for higher $k$ modes. The deviations from the GR results start to become visible for $k\geq k_{eq}\approx0.01$ Mpc$^{-1}$. These results were obtained by letting the initial conditions \eqref{pert_initcond} as free parameters and picking those that minimize the difference between the matter power spectra of $\Lambda$CDM and our model.

\begin{figure}[t]
	\includegraphics[width=\textwidth]{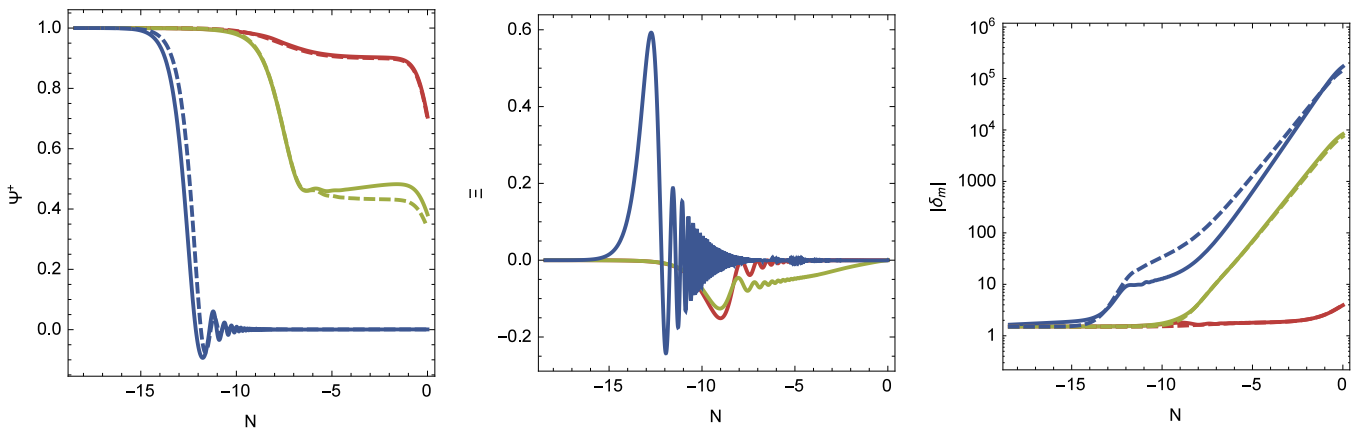}
	\caption{\label{RadMat_Pert_Evol}The evolution of the perturbation variables: $\Psi^+$ (left); $\Xi$ (middle); $\delta_m$ (right); since the radiation dominated epoch until the present time for different values of $k$: $k=2\times10^{-4}$ Mpc$^{-1}$ (red); $k=2\times10^{-2}$ Mpc$^{-1}$ (green); $k=2$ Mpc$^{-1}$ (blue). The evolution of the same variables in the $\Lambda$CDM model is plotted in dashed lines. Since the perturbation $\Xi$ vanishes in GR, the dashed lines do not appear on the middle panel. The $f(R)$ effective energy density accounts for the dark radiation content of the universe (the second model introduced in Sec.~\ref{Sec3B}).}
\end{figure}
\begin{figure}[t]
	\includegraphics[width=\textwidth]{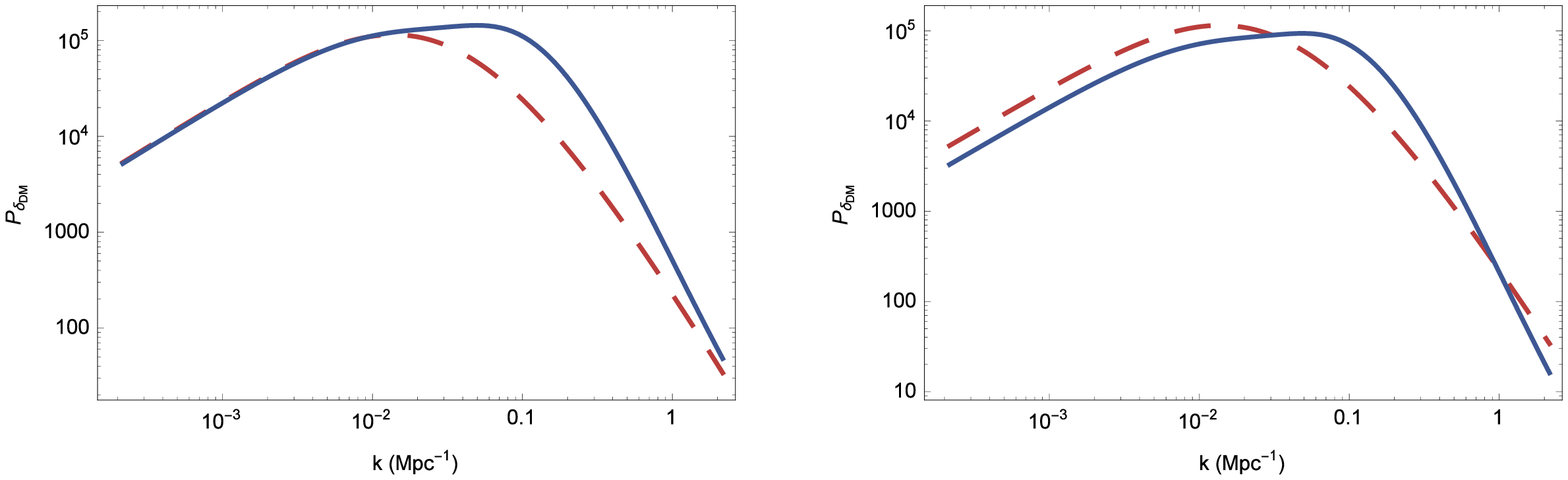}
	\caption{\label{RadMat_spectra}The matter power spectrum $\mathcal P_{\delta_m}$ for the $f(R)$ model that accounts for the dark radiation content of the universe (blue) and in the $\Lambda$CDM model (red discontinuous). On the left the matter power spectrum obtained for the initial conditions on Eq.~\eqref{pert_initcond}. On the right the matter power spectrum obtained by fine-tuning the initial conditions in order to minimize the deviations from the result of the $\Lambda$CDM model.}
\end{figure}

We next analyse the second model introduced in Sec.~\ref{Sec3B}, where now the $f(R)$ effective energy density again mimics a mGCG, of the type of Eq.~\eqref{mGCG_of_a} with $\beta=1/3$, even though it only accounts for a possible dark radiation content of the universe, as well as dark energy.
As before we evolve the perturbation variables since the early radiation dominated epoch until the present time, using the initial conditions
\begin{align}
	\Psi^+(N_{ini})=1,
	~~~~~~
	\left(\Psi^+\right)_N(N_{ini})=0,
	~~~~~~
	\Xi(N_{ini})=0,
	~~~~~~
	\left(\Xi\right)_N(N_{ini})=0.
\end{align}
In addition, we consider initial adiabatic conditions for the matter components, so that
\begin{align}
	\delta_m(N_{ini})= \frac{\delta_r(N_{ini})}{1+w_r},
	~~~~~~
	v_m(N_{ini}) = v_r(N_{ini}),
\end{align}
as it is commonly used in GR \cite{Dodelson:2003,Amendola:2010}. Notice that now the total matter perturbation $\delta$, and not $\delta_m$, is given by Eq.~\eqref{init_delta_val}. If we now apply the GR limit to the equation, taking into account the adiabaticity condition, we obtain the same result as in GR $\delta_m=-(3/2)\Psi^+$.
We find that when we consider radiation, due to the photons and the neutrinos components, the deviations of the function $f(R)$ from the Einstein-Hilbert action become much smaller, and, as a consequence, the evolution of the perturbation variables resembles more closely that of the $\Lambda$CDM model. In particular, for small comoving wave-numbers, $k$, the effects of the $f(R)$ gravity are negligible. This can be seen in Fig.~\ref{RadMat_Pert_Evol}, where we present the evolution of the perturbation variables $\Psi^+$, $\Xi$, and $\delta_m$, for different values of $k$. On the panel of the left-hand-side of Fig.~\ref{RadMat_spectra}, we plot the matter power spectrum obtained and compare it with the one of the $\Lambda$CDM model. As expected, the $f(R)$ gravity affects only the modes with higher $k$ \cite{Pogosian:2007sw,delaCruzDombriz:2008cp}, as the deviations from the $\Lambda$CDM spectrum become substantial for $k> k_{eq}\approx0.01$ Mpc$^{-1}$. On the panel on the rhs of Fig.~\ref{RadMat_spectra}, we show the matter power spectrum obtained by fine-tuning the initial conditions in order to minimize the deviations from the spectrum of the $\Lambda$CDM model, following the same methodology as before. We find that this procedure does not improve the results greatly, in particular it cannot resolve the difference in shape between the two power spectra for $k> k_{eq}\approx0.01$ Mpc$^{-1}$.

In this section we have evolved the first order metric and matter perturbations since the radiation dominated epoch until the present time and obtained the theoretical power spectrum of the dark matter perturbation. The numerical method employed was based on the integration of a set of coupled linear first order differential equations (four equation in the case of the first model presented in Sec.~\ref{Sec3B} and six equations in the case of the second model) and did not make use of any simplification of the equations involved, in particular the quasi static approximation \cite{delaCruzDombriz:2008cp}. Our numerical results show that the effects of the $f(R)$ corrections on the matter perturbation are stronger on the modes with higher wave-numbers, as is expected from the corrections to the closed evolution equation of $\delta_m$, derived in \cite{delaCruzDombriz:2008cp}. More precisely, this can be seen directly on the matter power spectra obtained: on the low $k$ end of the spectrum we find that the shape of the spectra obtained is consistent with the one of the $\Lambda$CDM model (up to a multiplicative factor); for modes with $k> k_{eq}\approx0.01$ Mpc$^{-1}$ the shape of the spectra obtained starts to diverge from the one in the $\Lambda$CDM model. Notice however that the matter power spectrum was obtained assuming that the initial amplitude profile of the initial conditions is that of $\Psi^+$ for single field inflation
\begin{align}
	\mathcal P _{\Psi^+} = \langle |\Psi^+|^2\rangle = \frac{8\pi^2}{9k^3}A_s\left(\frac{k}{k_0}\right)^{n_s-1},
\end{align}
which could be modified due to $f(R)$ effects. Here, $A_s=2.143\times10^{-9}$ is the curvature $\mathcal R$ power spectrum at the pivot scale $k_0=0.05$ Mpc$^{-1}$ , $n_s=0.9681$ is the scalar spectrum index as obtained by the Planck mission \cite{Ade:2015xua}, and we have used the relation $\mathcal{R}=(3/2)\Psi^+$ during the radiation epoch and for modes outside of the horizon.
Therefore, at the end of the numerical evolution of the perturbation variables with initial conditions given by \eqref{pert_initcond} we multiplied the results by ${\mathcal P_{\Psi^+}}^{1/2}$. As a matter of consistency, we applied the same method when searching for the set of initial conditions that gave the least deviation from the $\Lambda$CDM matter power spectrum. However, a different early inflationary history, in particular one originated from $f(R)$, might give different initial profiles for the metric perturbations.

%
%

\section{Discussion and Conclusions}
\label{Sec6}

The observed accelerated behaviour of the Hubble flow is the big puzzle of modern cosmology. Its explanation, under the standard of {\it dark energy}, implies to find out cosmic fluids capable of both giving rise to the today observed accelerated expansion and of tracing back the universe history giving the possibility, at certain epoch, of structure formation. From a fundamental physics point of view, it is extremely difficult to find out some material components giving rise to dark energy behaviours. The same situation exists for dark matter: even if its effects are evident at astrophysical and cosmological scales, no fundamental particle has been detected, up to now, to account for this ingredient. 

A different approach relies on the fact that GR could be {\it extended} at infra-red and ultraviolet scales enclosing further curvature invariants into the Einstein-Hilbert action. Several unification schemes, as strings, Kaluza-Klein theories and so on, are consistent with effective actions where higher-order curvature terms emerges as interaction terms. In general, any approach aimed to formulate the quantum field theory on curved space-time gives rise to further higher-order curvature terms. From this perspective, it is reasonable to investigate if such curvature corrections could account for inflation at early epochs and dark energy at late epochs. The big issue is to join these early and late behaviours passing through a radiation-matter dominated era where structure formation is possible. 

In this paper, we discussed $f(R)$ gravity models in order to see if such models could account also for radiation as we observe today. In other words, we investigate the possibility that, in addition to the explanation of present cosmic speed up, also a possible (dark) radiation component could 
be addressed by $f(R)$ gravity. 
Our analysis has been based on the modelling of $f(R)$ gravity according to the matter density $\rho$. In particular, assuming a generalized Chaplygin gas as the fluid fuelling the universe, it is possible to account for the accelerated behaviour (dark energy) and a further contribution due to a sort of dark radiation. It is shown that within these models $f(R)$ modification cannot account simultaneously for dark radiation and dark matter. We consider two models where, apart from the $f(R)$ effective density, the universe is sourced by (i) dust-like matter that accounts for baryonic and dark matter; and (ii) dust-like matter, that accounts for baryonic and dark matter, and normal radiation that accounts for the energy density of photons and neutrinos.

Our issue has been to constraints the models by using the Planck Collaboration 2015 data within a suitable cosmographic approach, developed for $f(R)$ gravity, and to obtain the related matter power spectrum. 
We note that in the two models we analysed, the $f(R)$ functions obtained will give rise to future instabilities, of the Dolgov-Kawasaki type \cite{Dolgov:2003px,Faraoni:2006sy}, as both the second derivative $f_{RR}$ and the effective squared mass of the scalaron, ${m^2_{\textrm{eff}} = f_R /f_{RR} - 2f /f_R}$ become negative. However, our work can be seen more as a demonstration of the method we have used and developed, rather than as a definitive result. In fact the two main models analysed can be seen as toy models although they can describe  the cosmological evolution since the radiation epoch till the present time without presenting any instabilities.
The result has been that $f(R)$ gravity can only contribute minimally to the (dark) radiation in order to avoid departures from the observed matter power spectrum at the smallest scales. In the case of $f(R)$ corrections accounting for all the radiation content, Fig.~\ref{Mat_spectra}, the matter power spectrum is several orders of magnitude higher than the observed; in the case of $f(R)$ corrections accounting only for the dark part of radiation, Fig.~\ref{RadMat_spectra}, which constitutes less than $1.4\%$ of the total radiation energy density, the matter power spectrum starts to diverge for scales of the order $0.01$~Mpc$^{-1}$, that is those scales that exit the horizon at the radiation dominated epoch. This fact, in our opinion, can be seen as a reliable constraint on viable $f(R)$ models.

%
%

\acknowledgments

The Authors are grateful to A. de la Cruz-Dombriz, R. Lazkoz, D. Sa\'ez-Gomez, V. Salzano, S. Tsujikawa, and D. Wands for enlightening discussions on the cosmological perturbations, the cosmography, and the matter power spectrum.
J.M. is thankful to UPV/EHU for a PhD fellowship and UBI for hospitality during completion of part of this work and acknowledges the support from the Basque government Grant No. IT592-13 (Spain).
The work of M.B.L. is supported by the Portuguese Agency “Funda\c{c}\~ao para a Ci\^encia e Tecnologia” through an ``Investigador FCT'' Research contract, with reference IF/01442/2013/CP1196/CT0001. She also wishes to acknowledge the support from the Portuguese Grants PTDC/FIS/111032/2009 and UID/MAT/00212/2013 and the partial support from the Basque government Grant No. IT592-13 (Spain). SC acknowledges the support of {\it Istituto Nazionale di Fisica Nucleare}, ({\it iniziativa specifica} QGSKY).

%
%

\appendix

\section{}
\label{app_B}

The following analysis of the $R$ derivatives of the function $f$ is done in terms of the variable $x$. With this in mind, we use Eq.~\eqref{x_of_R} to express $f_R$ and $f_{RR}$ in terms of $f_x$ and $f_{xx}$ as
\begin{align}
	\label{fR_B=1/3}
	f_R 	=& \frac{dx}{dR}f_x = \frac{1}{4\rho_{dS}}\frac{1+\alpha}{\alpha}x^{\frac{1}{1+\alpha}}f_x,
\end{align}
\begin{align}
	\label{fRR_B=1/3}
	f_{RR} =& \left(\frac{dx}{dR}\right)^2 f_{xx} + \frac{d^2x}{dR^2}f_x
		= \frac{1}{16\rho_{dS}^2}\frac{1+\alpha}{\alpha^2} x^{\frac{2}{1+\alpha}}
		\left[
			(1+\alpha)f_{xx}
			+ x^{-1}f_x
		\right],
\end{align}
while the derivatives $f_x$ and $f_{xx}$ can easily be computed from the rules \cite{Abramowitz1965,Olver2010}
\begin{align}
	\label{deriva1_2F1}
	\frac{d^n}{dx^n}\textrm{F}[b,c;d,x] = \frac{(b)_n(c)_n}{(d)_n}{}_2\textrm{F}_1[b+n,c+n;d+n,x],
\end{align}
\begin{align}
	\label{deriv2_2F1}
	\frac{d^n}{dx^n}\left(x^{d-1}\textrm{F}[b,c;d,x]\right) = (d-n)_n x^{d-n-1}\textrm{F}[b,c;d-n,x].
\end{align}
Here, $(y)_n$ is the Pochhammer symbol which denotes the rising factorial $(y)_n=y(y+1)...(y+n-1)$ \cite{Abramowitz1965,Olver2010}.

From the previous results, we deduce the expressions for $f_{1R}(x)$, $f_{1RR}(x)$, $f_{2R}(x)$ and $f_{2RR}(x)$
\begin{align}
	\label{f1R_B=1/3}
	f_{1R}(x)
	=& -\frac{1}{10\rho_{dS}}x^{\lambda_\alpha}
	\textrm{F}
	\left[
		1+\frac{1}{2}\lambda_\alpha,
		\lambda_\alpha;
		1+\frac{5}{4}\lambda_\alpha;
		x
	\right],
\end{align}
\begin{align}
	\label{f1RR_B=1/3}
	f_{1RR}(x)
		= -\frac{1}{20\alpha\rho_{dS}^2}x^{2\lambda_\alpha}
		&\Bigg\{
			\frac{3+2\alpha}{9+4\alpha}
			\textrm{F}
			\left[
				2+\frac{1}{2}\lambda_\alpha,
				1+\lambda_\alpha;
				2+\frac{5}{4}\lambda_\alpha;
				x
			\right]
			+ \frac{1}{2} x^{-1}
			\textrm{F}
			\left[
				1+\frac{1}{2}\lambda_\alpha,
				\lambda_\alpha;
				1+\frac{5}{4}\lambda_\alpha;
				x
			\right]
		\Bigg\},
\end{align}
\begin{align}
	\label{f2R_B=1/3}
	f_{2R}(x)
	=& -\frac{1-4\alpha}{16\alpha\rho_{dS}}x^{-\frac{1}{4}\lambda_\alpha}
	\textrm{F}
	\left[
		-\frac{1}{4}\lambda_\alpha,
		1-\frac{3}{4}\lambda_\alpha;
		1-\frac{5}{4}\lambda_\alpha;
		x
	\right],
\end{align}
\begin{align}
	\label{f2RR_B=1/3}
	f_{2RR}(x)
		= -\frac{1-4\alpha}{64\alpha^2\rho_{dS}^2}
		x^{\frac{3}{4}\lambda_\alpha -1}
		&\Bigg\{
			-\frac{5}{4}\textrm{F}
			\left[
				-\frac{1}{4}\lambda_\alpha,
				1-\frac{3}{4}\lambda_\alpha;
				-\frac{5}{4}\lambda_\alpha;
				x
			\right]
			+ \textrm{F}
			\left[
				-\frac{1}{4}\lambda_\alpha,
				1-\frac{3}{4}\lambda_\alpha;
				1-\frac{5}{4}\lambda_\alpha;
				x
			\right]
		\Bigg\}.
\end{align}
where $\lambda_\alpha\equiv1/(1+\alpha)$.


\section{}
\label{app:cosmography}

In a universe filled with radiation, $\rho_r$, dust like matter, $\rho_m$, and a dark-energy fluid, $\rho_{de}$, we can write the Friedmannn equation, in the context of GR, as
\begin{align}
	3H^2 = \rho_{m}+\rho_{r}+\rho_{de}.
\end{align}
In addition, if we can write $\rho_{de}=\rho_{de}(a)$ then $w_{de}\equiv p_{de}/\rho_{de}=w_{de}(a)$.
Using these results, and iteratively differentiating the Friedmannn equation, we can write the cosmographic parameters $q_0$, $j_0$, $s_0$ $l_0$ in terms of $\Omega_{r,0}$ and $\Omega_{m,0}$ and of the present day values of $w_{de}$ and its cosmic derivatives. In fact,
\begin{align}
	q_0 &= \frac{1+3w_{de,0}}{2}
		- \frac{3w_{de,0}}{2}\Omega_{m,0}
		+ \frac{1-3w_{de,0}}{2}\Omega_{r,0},
	\\
	j_0 &= \left[1 + \frac{9}{2}\left(1+w_{de,0}\right)w_{de,0} - \frac{3}{2}\left(\frac{\partial w_{de}}{\partial a}\right)_0\right]
	\nonumber\\&
		-\left[\frac{9}{2}\left(1+w_{de,0}\right)w_{de,0} - \frac{3}{2}\left(\frac{\partial w_{de}}{\partial a}\right)_0\right]\Omega_{m,0}
		- \left[\frac{9}{2}\left(w_{de,0}-\frac{1}{3}\right)\left(w_{de,0}+\frac{4}{3}\right) - \frac{3}{2}\left(\frac{\partial w_{de}}{\partial a}\right)_0\right]\Omega_{r,0},
	\\
	s_0 &= \frac{1}{4}\left[
		-\left(1+3w_{de,0}\right)\left(2+3w_{de,0}\right)\left(7+9w_{de,0}\right)
		+ 3\left(11+21w_{de,0}\right)\left(\frac{\partial w_{de}}{\partial a}\right)_0
		- 6 \left(\frac{\partial^2 w_{de}}{\partial a^2}\right)_0
	\right]
	\nonumber\\&
	+
		\frac{4}{3}\left[3w_{de,0}\left(9+19w_{de,0}+12w_{de,0}^2\right)
		-\left(11+24w_{de,0}\right)\left(\frac{\partial w_{de}}{\partial a}\right)_0
		+2\left(\frac{\partial^2 w_{de}}{\partial a^2}\right)_0\right]\Omega_{m,0}
	\nonumber\\&
		+\frac{8}{3}\left[
			-7 + w_{de,0}\left(2+3w_{de,0}\right)\left(5+6w_{de,0}\right)
			-\left(5+12w_{de,0}\right)\left(\frac{\partial w_{de}}{\partial a}\right)_0
			+\left(\frac{\partial^2 w_{de}}{\partial a^2}\right)_0
		\right]\Omega_{r,0}
	\nonumber\\&
		-\frac{9}{4}w_{de,0}\left[
			3\left(1+w_{de,0}\right) 
			- \left(\frac{\partial w_{de}}{\partial a}\right)_0
		\right]\Omega_{m,0}^2
	\nonumber\\&
		-\frac{3}{4}\left[
			18w_{de,0}\left(w_{de,0}+\frac{7}{6}\right)\left(w_{de,0}-\frac{1}{3}\right)
			+\left(1-6w_{de,0}\right)\left(\frac{\partial w_{de}}{\partial a}\right)_0
		\right]\Omega_{r,0}\Omega_{m,0}
	\nonumber\\&
		+\frac{3}{4}\left(1-3w_{de,0}\right)\left[
			-3\left(w_{de,0}+\frac{4}{3}\right)\left(w_{de,0}-\frac{1}{3}\right)
			+\left(\frac{\partial w_{de}}{\partial a}\right)_0
		\right]\Omega_{r,0}^2,
\end{align}
\begin{align}
	l_0 &= \frac{1}{4}\left[
		\left(1+3w_{de,0}\right)\left(2+3w_{de,0}\right)\left(5+6w_{de,0}\right)\left(7+9w_{de,0}\right)
		-3\left(
			71 + 246w_{de,0} + 207w_{de,0}^2
		\right)\left(\frac{\partial w_{de}}{\partial a}\right)_0
	\right.
	\nonumber\\&~~~~~~
	\left.
		+63\left(\frac{\partial w_{de}}{\partial a}\right)_0^2
		+\left(51+99w_{de,0}\right)\left(\frac{\partial^2 w_{de}}{\partial a^2}\right)_0
		-6\left(\frac{\partial^3 w_{de}}{\partial a^3}\right)_0
	\right]
	\nonumber\\&
		-\frac{3}{4}\bigg[
			\left(163 + 528w_{de,0}+639w_{de,0}^2+270w_{de,0}^3\right)w_{de,0}
			-\left(17 + 309w_{de,0} + 306w_{de,0}^2\right)\left(\frac{\partial w_{de}}{\partial a}\right)_0
			+24\left(\frac{\partial w_{de}}{\partial a}\right)_0^2
		\bigg.
	\nonumber\\&~~~~~~
		\bigg.
			+\left(17+28w_{de,0}\right)\left(\frac{\partial^2 w_{de}}{\partial a^2}\right)_0
			-2\left(\frac{\partial^3 w_{de}}{\partial a^3}\right)_0
		\bigg]\Omega_{m,0}
	\nonumber\\&
		-\frac{1}{2}\bigg[
			-\left(1-3w_{de,0}\right)\left(140+342w_{de,0}+251w_{de,0}^2+135w_{de,0}^3\right)
			-9\left(8+47w_{de,0}+51w_{de,0}^2\right)\left(\frac{\partial w_{de}}{\partial a}\right)_0
		\bigg.
	\nonumber\\&~~~~~~
		\left.
			+36\left(\frac{\partial w_{de,0}}{\partial a}\right)_0^2 
			+ 21\left(1+3w_{de,0}\right)\left(\frac{\partial^2 w_{de}}{\partial a^2}\right)_0
			-3\left(\frac{\partial^3 w_{de}}{\partial a^3}\right)_0
		\right]\Omega_{r,0}
	\nonumber\\&
	+
		\frac{1}{4}\left[
			7+12w_{de,0}+6w_{de,0}^2
			-3w_{de,0}\left(7+11w_{de}\right)\left(\frac{\partial w_{de}}{\partial a}\right)_0
			+\left(\frac{\partial w_{de}}{\partial a}\right)_0^2
			+3w_{de,0}\left(\frac{\partial^2 w_{de}}{\partial a^2}\right)_0
		\right]\Omega_{m,0}^2
	\nonumber\\&
		+\frac{3}{4}\left[
			-3w_{de,0}\left(1-3w_{de,0}\right)\left(37+53w_{de,0}+24w_{de,0}^2\right)
			+\left(23-99w_{de,0}-198w_{de,0}^2\right)\left(\frac{\partial w_{de}}{\partial a}\right)_0
		\right.
	\nonumber\\&~~~~~~
		\left.
			+6\left(\frac{\partial w_{de}}{\partial a}\right)_0^2
			-3\left(1-6w_{de,0}\right)\left(\frac{\partial^2 w_{de}}{\partial a^2}\right)_0
		\right]\Omega_{r,0}\Omega_{m,0}
	\nonumber\\&
		+\frac{1}{4}\left[
			\left(1-3w_{de,0}\right)^2\left(70+87w_{de,0}+36w_{de,0}^2\right)
			+3\left(1-3w_{de,0}\right)\left(23+33w_{de,0}\right)\left(\frac{\partial w_{de}}{\partial a}\right)_0
		\right.
	\nonumber\\&~~~~~~
		\left.
			+9\left(\frac{\partial w_{de}}{\partial a}\right)_0^2
			-9\left(1-3w_{de,0}\right)\left(\frac{\partial^2 w_{de}}{\partial a^2}\right)_0
		\right]\Omega_{r,0}^2.
\end{align}
Notice that these expressions generalize previous analysis (see for example Ref.~\cite{Capozziello:2011tj}) as they are valid for any equation of state parametrized exclusively by the scale factor and include explicitly the contribution of radiation. 

\bibliography{fr-chaplygin05_08_2015.bbl}
 
\end{document}